\documentclass[aps,showpacs,preprintnumbers,amsmath,amssymb,twocolumn,superscriptaddress,prx,floatfix,nofootbib]{revtex4-2}

\usepackage{amsfonts}
\usepackage{amsmath}
\usepackage{amssymb}
\usepackage{graphicx}
\usepackage{dcolumn}
\usepackage{hyperref}
\usepackage{bm}
\usepackage{xkcdcolors}
\usepackage{tabularx}
\usepackage{epstopdf}
\usepackage{xcolor}
\usepackage{mathrsfs}
\usepackage{caption}
\usepackage{subcaption}
\usepackage{mathtools}
\usepackage{float}
\usepackage{algpseudocode}
\usepackage{verbatim}
\usepackage{comment}
\usepackage{cleveref}
\usepackage{ragged2e}
\usepackage{physics}
\usepackage{nccmath}
\DeclareCaptionJustification{justified}{\justifying}
\newcolumntype{C}[1]{>{\centering\arraybackslash}p{#1}}

\begin{document}

\title{Reconciling econometrics with continuous maximum-entropy network models}

\author{Marzio Di Vece}
\email{marzio.divece@imtlucca.it}
\affiliation{IMT School for Advanced Studies, P.zza San Francesco 19, 55100 Lucca (Italy)}
\author{Diego Garlaschelli}
\affiliation{IMT School for Advanced Studies, P.zza San Francesco 19, 55100 Lucca (Italy)}
\affiliation{Lorentz Institute for Theoretical Physics, Leiden University, Niels Bohrweg 2, 2333CA Leiden (The Netherlands)}
\affiliation{INdAM-GNAMPA Istituto Nazionale di Alta Matematica (Italy)}
\author{Tiziano Squartini}
\affiliation{IMT School for Advanced Studies, P.zza San Francesco 19, 55100 Lucca (Italy)}
\affiliation{INdAM-GNAMPA Istituto Nazionale di Alta Matematica (Italy)}
\affiliation{Institute for Advanced Study (IAS), University of Amsterdam, Oude Turfmarkt 145, 1012GC Amsterdam (The Netherlands)}

\date{\today}

\begin{abstract}
In the study of economic networks, econometric approaches interpret the traditional Gravity Model specification as the expected link weight coming from a probability distribution whose functional form can be chosen arbitrarily, while statistical-physics approaches construct maximum-entropy distributions of weighted graphs, constrained to satisfy a given set of measurable network properties. In a recent, companion paper, we integrated the two approaches and applied them to the World Trade Web, i.e. the network of international trade among world countries. While the companion paper dealt only with discrete-valued link weights, the present paper extends the theoretical framework to continuous-valued link weights. In particular, we construct two broad classes of maximum-entropy models, namely the integrated and the conditional ones, defined by different criteria to derive and combine the probabilistic rules for placing links and loading them with weights. In the integrated models, both rules follow from a single, constrained optimization of the continuous Kullback-Leibler divergence; in the conditional models, the two rules are disentangled and the functional form of the weight distribution follows from a conditional, optimization procedure. After deriving the general functional form of the two classes, we turn each of them into a proper family of econometric models via a suitable identification of the econometric function relating the corresponding, expected link weights to macroeconomic factors. After testing the two classes of models on  World Trade Web data, we discuss their strengths and weaknesses.
\end{abstract}

\maketitle

\section{Introduction}

Over the last couple of decades, the growth of network science has impacted several disciplines by establishing new, empirical facts about many, real-world systems, in terms of the structural, network properties that are typically found in those systems. In the context of trade economics, the growing availability of data about import/export relationships among world countries has prompted researchers to explore and model the architecture of the international trade network, or World Trade Web (WTW)~\cite{Barigozzi2010,Fronczak2012a,Fronczak2012b,Serrano2003,Garlaschelli2004,Garlaschelli2005,Fagiolo2010,Fagiolo2008a,Fagiolo2008b,Schweitzer2009,Vitali2011,Schiavo2010}.

This approach has complemented, and in many ways enriched, the traditional econometric exercise of modelling individual trade flows, i.e. relating the volume of individual trade exchanges to the most relevant covariates (generally, macroeconomic factors) they may depend on. The earliest example of an econometric model for international trade is the celebrated Gravity Model (GM)~\cite{Tinbergen1962} that predicts that the expected value $\langle w_{ij}\rangle_\text{GM}$ of the trade volume $w_{ij}$ from country $i$ to country $j$ can be expressed via the econometric function

\begin{equation}\label{GMexpression}
\langle w_{ij}\rangle_\text{GM}=f(\omega_i,\omega_j,d_{ij}|\underline{\psi})=\tau\omega_i^{\beta_1}\omega_j^{\beta_2}d_{ij}^{\gamma}
\end{equation}
where $\omega_i\equiv{\text{GDP}_i}/{\overline{\text{GDP}}}$ is the GDP of country $i$ divided by the arithmetic mean of the GDPs of all countries, $d_{ij}$ is the geographic distance between (generally the capitals of) countries $i$ and $j$ and $\underline{\psi}\equiv(\tau,\beta_1,\beta_2,\gamma)$ is a vector of parameters (notice that $\beta_1\equiv\beta_2$ when the direction of the exchanges is disregarded, as we will, throughout the paper, and $\tau$ takes care of dimensional units). Equation~\ref{GMexpression} has a long tradition in successfully explaining the existing (i.e. \emph{positive}) trade volumes between pairs of countries. Outside economics, the gravity equation has also been extensively employed in studies concerning transportation, migration~\cite{Wilson1969} and the maximization of utility functions constrained to satisfy requirements on the rate of information acquisition~\cite{Wang2021}.

Although accurate in reproducing the positive trade volumes, the traditional GM cannot replicate structural network properties, unless the topology is completely fixed via a separate approach~\cite{Fagiolo2010b}. Indeed, if $\mathbf{W}$ denotes the weighted adjacency matrix of the WTW, where the entry $w_{ij}$ represents the trade volume from country $i$ to country $j$,  Eq.~\ref{GMexpression} predicts that the expected matrix $\langle\textbf{W}\rangle_\text{GM}$ has no off-diagonal zeroes, i.e. an expected positive trade relationship exists between all countries. This means that, when interpreted as an expected value of a regression with small, symmetric, zero-mean (e.g. Gaussian) noise, Eq.~\ref{GMexpression} predicts a fully connected network: if $a_{ij}$ denotes the generic entry of the binary adjacency matrix $\mathbf{A}=\Theta[\mathbf{W}]$ (equal to $a_{ij}=1$ if a positive trade volume from country $i$ to country $j$ is present, i.e. $w_{ij}>0$, and equal to $a_{ij}=0$ if a zero trade is, instead, observed, i.e. $w_{ij}=0$), then Eq.~\ref{GMexpression} predicts almost surely $a_{ij}=1$, $\forall\:i\neq j$. This result is in obvious contrast with empirical data, which show that the WTW has a rich topological architecture, characterized by a broad degree distribution, (dis)assortative and clustering patterns and other properties~\cite{Barigozzi2010,Fronczak2012a,Fronczak2012b,Serrano2003,Garlaschelli2004,Garlaschelli2005,Fagiolo2010,Fagiolo2008a,Fagiolo2008b,Schweitzer2009,Vitali2011,Schiavo2010}.

In order to overcome such a limitation, the plain Gravity Model needs to be `dressed' with a probability distribution $Q(\mathbf{W})$ that produces $\langle\textbf{W}\rangle_\text{GM}$ as the expected value while, at the same time, accounts for null outcomes as well (i.e. those entries reading $w_{ij}=0$ and representing missing links in the network)~\cite{Helpman2008}. Clearly, the support $\mathbb{W}$ of the probability $Q(\mathbf{W})$ should not include matrices with negative numbers. From the Sixties on, the GM has indeed been interpreted as the expected value of a probability distribution whose functional form needs to be determined.

Trade econometrics models the tendency of countries to establish trade relationships relating it to accepted, macroeconomic determinants (the so-called `factors') such as GDP and geographic distance, as in the expression of Eq.~\ref{GMexpression}. Econometricians have considered increasingly flexible distributions, the most recent versions of them being capable of disentangling the estimation of the presence of a trade exchange from the estimation of the traded amount. This has led to the definition of two distinct classes of models, i.e. \emph{zero-inflated} models~\cite{Burger2009} and \emph{hurdle} models~\cite{Long1997}. Zero-inflated (ZI) models have been introduced to model the following two scenarios: the possibility of exchanging a zero amount of goods even after having established a trade partnership (e.g. because of a limited trading capacity); the possibility of establishing a very small amount of trade (in fact, so small to be compatible with statistical noise and, as such, removed). A general drawback of employing ZI models is that of predicting sparser-than-observed network structures~\cite{Marzio2022}; moreover, only discrete distributions (specifically, either the Poisson or the negative-binomial one~\cite{Burger2009}) have been considered, so far, to carry out the proper weight-estimation step. Hurdle models, introduced to overcome the limitations affecting zero-inflated models, can predict zeros only at the first step~\cite{Long1997}: in any case, the presence of links is established by employing either a logit or a probit estimation step.

Network science has tackled the aforementioned inference problem using techniques rooted in statistical physics. The most prominent examples descend from the Maximum-Entropy Principle (MEP)~\cite{Jaynes1957a,Jaynes1957b,Jaynes1982} applied to network ensembles~\cite{DSBook,Cimini2019}, prescribing to maximize Shannon entropy~\cite{Shannon,Cover} in a constrained fashion to obtain the maximally unbiased distribution of networks compatible with a chosen set of structural constraints. This approach is formally equivalent to the construction of so-called Exponential Random Graphs (ERGs) for social network analysis~\cite{Wasserman1994} but differs in the typical choice of the constraints: in particular, when the enforced constraints are local, such as the degree (number of links) and the strength (total link weight) of each node, maximum-entropy network models have been shown to successfully replicate both the topology and the weights of many economic and financial networks, including the WTW~\cite{Garlaschelli2004,Garlaschelli2005,Squartini2011a,Squartini2011b,Mastrandrea2014b,Squartini2014,Almog2015,Almog2017,Almog2019,Marzio2022}. The entire framework can also  accommodate possibly degenerate, discrete-valued, single- or multi-edges \cite{Sagarra2015}.

Although maximum-entropy models have been also studied from an economic perspective (see~\cite{Bargigli2014} for a discussion of the economic relevance of the constraints defining the Poisson and the geometric network models), it is only recently that progress has been made to reconcile the above two approaches, allowing for economic factors parametrizing the maximum-entropy probability distribution producing links and weights~\cite{Garlaschelli2004,Garlaschelli2005,Squartini2014,Almog2015,Almog2017,Almog2019,Marzio2022} or by introducing network-related statistics into otherwise purely econometric models~\cite{Herman2022}. On one hand, the novel framework enriches the methods developed by network scientists with an econometric interpretation; on the other, it enlarges the list of candidate distributions usable for econometric purposes.

With this contribution, we refine the theoretical picture provided in a companion paper~\cite{Marzio2022}, introducing models to infer the topology and the weights of undirected networks defined by continuous-valued data. In order to do so, we present a theoretical, physics-inspired framework capable of accommodating both integrated and conditional, continuous models, our goal being threefold: 1) testing the performance of both classes of models on the WTW in order to understand which one is best suited for the task; 2) offering a principled derivation of currently available, conditional, econometric models; 3) enlarging the list of continuous-valued distributions to be used for econometric purposes. From an econometric point of view, our work moves along the methodological guidelines defining the class of Generalized Linear Models (GLMs)~\cite{Nelder1972} while enriching it with distributions defined by both econometric and structural parameters. From a statistical physics point of view, our work expands the class of maximum-entropy network models~\cite{DSBook} or weighted ERGs~\cite{Wasserman1994} and endows them with macroeconomic factors replacing certain model parameters.\\

The rest of the paper is organized as follows: in Sec.~\ref{sec:conditional}, after introducing the basic quantities, we derive the class of conditional models; in Sec.~\ref{sec:integrated} we derive the class of integrated models; in Sec.~\ref{sec:results} we  apply all models to the analysis of WTW data; in Sec.~\ref{sec:discussion} we discuss the results and provide our concluding remarks.

\section{Conditional Models\label{sec:conditional}}

Discrete maximum-entropy models can be derived by performing a constrained maximization of Shannon entropy~\cite{Jaynes1957a,Jaynes1957b,Jaynes1982}. However, unlike the companion paper~\cite{Marzio2022}, our focus, here, is on continuous probability distributions. In such a case, mathematical problems are known to affect the definition of Shannon entropy and the resulting inference procedure. To restore the framework, one has to introduce the Kullback-Leibler (KL) divergence $D_\text{KL}(Q||R)$ of a distribution $Q$ from a prior distribution $R$ and re-interpret the maximization of the entropy of $Q$ as the minimization of $D_\text{KL}(Q||R)$ from a given prior distribution $R$. In formulas, the KL divergence is defined as

\begin{equation}
D_\text{KL}(Q||R)=\int_\mathbb{W}Q(\mathbf{W})\ln \frac{Q(\mathbf{W})}{R(\mathbf{W})}d\mathbf{W}
\end{equation}
where $\mathbf{W}$ is one of the possible values of a continuous random variable (in our setting, an entire network with continuous-valued link weights), $\mathbb{W}$ is the set of possible values that $\mathbf{W}$  can take, $Q(\mathbf{W})$ is the (multivariate) probability density function to be estimated and $R(\mathbf{W})$ plays the role of prior  distribution, the divergence of $Q(\mathbf{W})$ from which must be minimized. Such an optimization scheme embodies the so-called \emph{Minimum Discrimination Information Principle} (MDIP), originally proposed by Kullback and Leibler~\cite{Kullback1951} and implementing the idea that, given a prior distribution $R(\mathbf{W})$ and new information that becomes available, an updated distribution $Q(\mathbf{W})$ should be chosen in order to make its discrimination from $R(\mathbf{W})$ as hard as possible. In other words, the MDIP demands that new data produce an information gain that is as small as possible. The use of the KL divergence is widespread in the fields of information theory~\cite{Shannon} and machine learning~\cite{Goodfellow2014}, e.g. as a loss function within the Generative Adversarial Network (GAN) scheme (the aim of the `generating' neural network being that of producing samples that cannot be distinguished from those constituting the training set by the `discriminating' neural network).

In order to introduce the class of conditional models, we write the posterior distribution $Q(\mathbf{W})$ as

\begin{equation}
Q(\mathbf{W})=P(\mathbf{A})Q(\mathbf{W}|\mathbf{A}),
\end{equation}
where $\mathbf{A}$ denotes the adjacency matrix for the binary projection of the weighted network $\mathbf{W}$. The above equation allows us to split the KL divergence into the following sum of three terms

\begin{equation}
D_\text{KL}(Q||R)=S(Q,R)-S(P)-S(Q_\bot|P)
\end{equation}
where

\begin{equation}
S(P)=-\sum_{\mathbf{A}\in\mathbb{A}}P(\mathbf{A})\ln P(\mathbf{A})
\end{equation}
is the \emph{Shannon entropy} of the probability distribution describing the binary projection of the network structure,

\begin{equation}
S(Q_\bot|P)=-\sum_{\mathbf{A}\in\mathbb{A}}P(\mathbf{A})\int_{\mathbb{W}_\mathbf{A}}Q(\mathbf{W}|\mathbf{A})\ln Q(\mathbf{W}|\mathbf{A})d\mathbf{W}
\end{equation}
is the \emph{conditional Shannon entropy} of the probability distribution of the weighted network structure given the binary projection and

\begin{equation}
S(Q,R)=-\sum_{\mathbf{A}\in\mathbb{A}}P(\mathbf{A})\int_{\mathbb{W}_\mathbf{A}}Q(\mathbf{W}|\mathbf{A})\ln R(\mathbf{W})d\mathbf{W}
\end{equation}
is the \emph{cross entropy} quantifying the amount of information required to identify a weighted network sampled from the distribution $Q(\mathbf{W})$ by employing the distribution $R(\mathbf{W})$. When continuous models are considered, $S(Q_\bot|P)$ is defined by a first sum running over all the binary configurations within the ensemble $\mathbb{A}$ and an integral over all the weighted configurations that are compatible with each, specific, binary structure - embodied by the adjacency matrix $\mathbf{A}$, i.e. such that $\mathbb{W}_\mathbf{A}=\{\mathbf{W}:\Theta[\mathbf{W}]=\mathbf{A}\}$).

The expression for $S(Q,R)$ can be further manipulated as follows. Upon separating the prior distribution itself into a purely binary part and a conditional, weighted one, one can write

\begin{equation}
R(\mathbf{W})=T(\mathbf{A})R(\mathbf{W}|\mathbf{A})
\end{equation}
an expression that allows us to write $S(Q,R)$ as

\begin{align}
S(Q,R)=&-\sum_{\mathbf{A}\in\mathbb{A}}P(\mathbf{A})\ln T(\mathbf{A})\nonumber\\
&-\sum_{\mathbf{A}\in\mathbb{A}}P(\mathbf{A})\int_{\mathbb{W}_\mathbf{A}}Q(\mathbf{W}|\mathbf{A})\ln R(\mathbf{W}|\mathbf{A})d\mathbf{W}
\end{align}
which, in turn, allows the KL divergence to be rewritten as 

\begin{equation}
D_\text{KL}(Q||R)=-D_\text{KL}(P||T)-D_\text{KL}(Q_\bot||R_\bot)
\end{equation}
i.e. as a sum of two terms, one of which involves conditional distributions; specifically,

\begin{align}
D_\text{KL}(P||T)&=-\sum_{\mathbf{A}\in\mathbb{A}}P(\mathbf{A})\ln\frac{P(\mathbf{A})}{T(\mathbf{A})},\\
D_\text{KL}(Q_\bot||R_\bot)&=-\sum_{\mathbf{A}\in\mathbb{A}}P(\mathbf{A})\int_{\mathbb{W}_\mathbf{A}}Q(\mathbf{W}|\mathbf{A})\ln\frac{Q(\mathbf{W}|\mathbf{A})}{R(\mathbf{W}|\mathbf{A})}d\mathbf{W}
\end{align}
with $T(\mathbf{A})$ representing the binary prior and $R(\mathbf{W}|\mathbf{A})$ representing the conditional, weighted one. In what follows, we will deal with completely uninformative priors: this amounts at considering the somehow `simplified' expression

\begin{equation}
D_\text{KL}(Q||R)=-S(P)-S(Q_\bot|P)
\end{equation}
with

\begin{align}
S(P)&=-\sum_{\mathbf{A}\in\mathbb{A}}P(\mathbf{A})\ln P(\mathbf{A}),\\
S(Q_\bot|P)&=-\sum_{\mathbf{A}\in\mathbb{A}}P(\mathbf{A})\int_{\mathbb{W}_\mathbf{A}}Q(\mathbf{W}|\mathbf{A})\ln Q(\mathbf{W}|\mathbf{A})d\mathbf{W}.
\end{align}

The (independent) constrained optimization of $S(P)$ and $S(Q_\bot|P)$ represents the starting point for deriving the members of the class of conditional models.

\subsection{Choosing the binary constraints}\label{choosingbinaryconditional}

The functional form controlling for the binary part of conditional models can be derived by carrying out a constrained maximization of the binary Shannon entropy

\begin{equation}
S(P)=-\sum_{\mathbf{A}\in\mathbb{A}}P(\mathbf{A})\ln P(\mathbf{A})
\end{equation}
leading to a probability mass function reading

\begin{equation}
P(\mathbf{A})=\frac{e^{-H(\mathbf{A})}}{\sum_\mathbb{A}e^{-H(\mathbf{A})}}
\end{equation}
where the functional form of the Hamiltonian reads $H(\mathbf{A})=\sum_{i<j}\alpha_{ij} a_{ij}$. This choice induces a factorization of the probability mass function $P(\mathbf{A})$, which becomes

\begin{equation}\label{topconditionalL}
P(\mathbf{A})=\prod_{i<j}p_{ij}^{a_{ij}}(1-p_{ij})^{1-a_{ij}}
\end{equation}
i.e. a product of a number of Bernoulli-like probability mass functions with $p_{ij}=\frac{x_{ij}}{1+x_{ij}}$, $\forall\:i<j$, where $x_{ij}=e^{-\alpha_{ij}}$ is the Lagrange multiplier controlling for the generic entry of the adjacency matrix $\mathbf{A}$.

In what follows, we will consider the specification of the tensor-like Hamiltonian introduced above reading $\alpha_{ij}=\alpha_i+\alpha_j$, $\forall\:i<j$, a choice inducing the Undirected Binary Configuration Model (UBCM), characterized by the following, pair-specific probability coefficient

\begin{equation}\label{BCMpij}
p_{ij}^\text{UBCM}=\frac{x_ix_j}{1+x_ix_j}
\end{equation}
and ensuring the entire degree sequence of the network at hand to be reproduced.

The econometric reparametrization of the UBCM can be achieved by posing $x_i\equiv\sqrt{\delta}\omega_i$, $\forall\:i$, a choice inducing the so-called fitness model (FM), characterized by the pair-specific probability coefficient

\begin{equation}\label{Fitnesspij}
p_{ij}^\text{FM}=\frac{\delta\omega_i\omega_j}{1+\delta\omega_i\omega_j}
\end{equation}
and requiring the estimation of a global parameter only, i.e. $\delta$. The FM represents a particular case of the logit model \cite{Walker1967}, being defined by a vector of external properties (the `fitnesses') that replace the information provided by some kind of (otherwise) purely structural properties~\cite{Caldarelli2003,Garlaschelli2004}: in fact, Eq.~\ref{Fitnesspij} can be equivalently rewritten as $\text{logit}\left[p_{ij}^\text{FM}\right]\equiv e^{\underline{X}\cdot\underline{\phi}}$ with $\underline{X}\equiv[1,\ln(\omega_i\omega_j)]$ and $\underline{\phi}\equiv[\ln\delta,1]$. The global constant $\delta$ can be determined by imposing the total number of links as the only constraint. Remarkably, the fitness model has been proven to reproduce the (binary) properties of a wide spectrum of real-world systems \cite{Squartini2011a, Garlaschelli2005} as accurately as the UBCM, although requiring much less information.

In what follows, we will consider both the UBCM and the FM specifications.

\subsection{Choosing the weighted constraints}

The constrained maximization of $S(Q_\bot|P)$ proceeds by specifying the following set of weighted constraints

\begin{align}
1&=\int_{\mathbb{W}_\mathbf{A}}P(\mathbf{W}|\mathbf{A})d\mathbf{W},\:\forall\:\mathbf{A}\in\mathbb{A}\\
\langle C_\alpha\rangle&=\sum_{\mathbf{A}\in\mathbb{A}}P(\mathbf{A})\int_{\mathbb{W}_\mathbf{A}}Q(\mathbf{W}|\mathbf{A})C_\alpha(\mathbf{W})d\mathbf{W},\:\forall\:\alpha
\end{align}
the first condition ensuring the normalization of the probability distribution and the vector $\{C_\alpha(\mathbf{W})\}$ representing the `proper' set of weighted constraints (weights are, now, treated as continuous random variables, i.e. $w_{ij}\in\mathbb{R}^+_0$, $\forall\:i<j$). They induce the distribution reading

\begin{equation}
Q(\mathbf{W}|\mathbf{A})=
\begin{cases}
\frac{e^{-H(\mathbf{W})}}{Z_\mathbf{A}}, & \:\mathbf{W}\in\mathbb{W}_\mathbf{A}\\
0, & \:\mathbf{W}\notin\mathbb{W}_\mathbf{A}
\end{cases}
\label{eq_25}
\end{equation}
where $H(\mathbf{W})=\sum_\alpha\psi_\alpha C_\alpha$ is the so-called Hamiltonian, listing the constrained quantities, and $Z_\mathbf{A}=\int_{\mathbb{W}_\mathbf{A}}e^{-H(\mathbf{W})}d\mathbf{W}$ is the partition function, conditional on the `fixed topology' $\mathbf{A}$.

The explicit functional form of $Q(\mathbf{W}|\mathbf{A})$ can be obtained only once the functional form of the constraints has been specified. In what follows, we will deal with the Hamiltonian reading

\begin{align}
H(\mathbf{W})=\sum_{i<j}f(w_{ij}|\beta_{ij},\xi_{ij},\gamma_{ij})
\end{align}
with the Lagrange multipliers $\left(\beta_{ij},\xi_{ij},\gamma_{ij}\right)$ satisfying the following requirements:

\begin{itemize}
\item $\beta_{ij}\equiv\beta_0+\beta_{ij}$, where $\beta_0$ is the Lagrange multiplier associated with the total weight $\sum_{i<j}w_{ij}\equiv W_1$ and $\beta_{ij}$ encodes the dependence on purely econometric quantities;

\item $\xi_{ij}$ will be kept either in its dyadic form, to constrain the logarithm of each weight, or in its global form, $\xi_{ij}\equiv\xi_0$, to constrain the sum of the logarithms of the weights, i.e. $\sum_{i<j}\ln(w_{ij})\equiv W_2$;

\item $\gamma_{ij}\equiv\gamma_0$ plays the role of the Lagrange multiplier associated with (a function of) the total variance of the logarithms of the weights, i.e. $\sum_{i<j}\ln^2(w_{ij})\equiv W_3$.
\end{itemize}

\subsubsection{Conditional exponential model.}

Let us start by considering the simplest, conditional model, defined by the positions $\gamma_{ij}=\xi_{ij}=0$ and inducing the Hamiltonian

\begin{align}
H(\mathbf{W})&=\sum_{i<j}f(w_{ij}|\beta_0+\beta_{ij})\nonumber\\
&=\sum_{i<j}(\beta_0+\beta_{ij})w_{ij};
\end{align}
inserting the expression above into Eq.~\ref{eq_25} leads to the distribution

\begin{align}\label{exponential_conditional_distribution}
Q(\mathbf{W}|\mathbf{A})&=\prod_{i<j}q_{ij}(w_{ij}|a_{ij})\nonumber\\
&=\prod_{i<j}\frac{e^{-(\beta_0 + \beta_{ij})w_{ij}}}{\zeta_{ij}}\nonumber\\
&=\prod_{i<j}(\beta_0 + \beta_{ij})e^{-(\beta_0 + \beta_{ij})w_{ij}}
\end{align}
and each node pair-specific distribution induces a (conditional) expected weight reading

\begin{equation}\label{cond_exp_w}
\langle w_{ij}|a_{ij}=1\rangle=\frac{1}{\beta_0+\beta_{ij}}.
\end{equation}

From a purely topological point of view, constraining each weight \emph{and} their total sum is redundant. However, this is no longer true when turning the conditional, exponential model into a proper econometric one. Its econometric reparametrization should be consistent with the literature on trade, stating that the weights are monotonically increasing functions of the gravity specification, i.e. $\langle w_{ij}\rangle_\text{GM}=e^{\rho+\beta\cdot\ln(\omega_i\omega_j)+\gamma\cdot\ln(d_{ij})}\equiv z_{ij}$, $\forall\:i<j$ and with $e^\rho\equiv\tau$; for this reason, the \emph{link function} usually associated with the exponential distribution prescribes to identify the \emph{linear predictor} with the inverse of the purely econometric parameter of the model, i.e.

\begin{equation}
\beta_{ij}\equiv z_{ij}^{-1},
\end{equation}
a position that turns Eq.~\ref{cond_exp_w} into

\begin{equation}\label{firstmoment_cond_exp}
\langle w_{ij}|a_{ij}=1\rangle=\frac{1}{\beta_0+z_{ij}^{-1}}=\frac{z_{ij}}{1+\beta_0 z_{ij}};
\end{equation}
notice that the only structural constraint is, now, represented by the total weight (see also Appendix A).

\subsubsection{Conditional gamma model.}

Let us, now, consider a different Hamiltonian, constraining each weight, their total sum and the sum of their logarithms, i.e.

\begin{align}
H(\mathbf{W})&=\sum_{i<j}f(w_{ij}|\beta_0+\beta_{ij},\xi_0)\nonumber\\
&=\sum_{i<j}[(\beta_0+\beta_{ij})w_{ij}+\xi_{0}\ln(w_{ij})]\nonumber\\
&=\sum_{i<j}\beta_{ij}w_{ij}+\beta_0W_1+\xi_{0}W_2;
\end{align}
it induces the distribution reading

\begin{align}
Q(\mathbf{W}|\mathbf{A})&=\prod_{i<j}q_{ij}(w_{ij}|a_{ij})\nonumber\\
&=\prod_{i<j}\dfrac{e^{-(\beta_0+\beta_{ij})w_{ij}-\xi_0\ln(w_{ij})}}{\zeta_{ij}}\nonumber\\
&=\prod_{i<j}\dfrac{e^{-(\beta_0+\beta_{ij})w_{ij}}w_{ij}^{-\xi_0}}{\zeta_{ij}}\nonumber\\
&=\prod_{i<j}\dfrac{(\beta_0+\beta_{ij})^{1-\xi_0}}{\Gamma(1-\xi_0)}e^{-(\beta_0+\beta_{ij})w_{ij}}w_{ij}^{-\xi_0};
\label{distribution_cgm}
\end{align}
each node pair-specific distribution is characterized by a (conditional) expected weight reading

\begin{equation}\label{w_cond_cgm}
\langle w_{ij}|a_{ij}=1\rangle=\frac{1-\xi_0}{\beta_0+\beta_{ij}}
\end{equation}
and by a (conditional) expected logarithmic weight reading

\begin{equation}\label{lnw_cond_cgm}
\langle\ln(w_{ij})|a_{ij}=1\rangle=\psi(1-\xi_0)-\ln(\beta_0+\beta_{ij})
\end{equation}
where the function $\psi(x)=\Gamma'(x)/\Gamma(x)$ is the so-called \emph{digamma function}.\\

Such a model can be turned into a proper, econometric one by considering the inference scheme of the gamma model with inverse response, which allows us to identify the linear predictor with the inverse of the purely econometric parameter of the model, i.e.

\begin{equation}
\beta_{ij}\equiv z_{ij}^{-1}
\end{equation}
a position that, in turn, leads to the expressions

\begin{equation}\label{firstmoment_cond_gamma}
\langle w_{ij}|a_{ij}=1\rangle=\frac{1-\xi_0}{\beta_0+z_{ij}^{-1}}=\frac{(1-\xi_0)z_{ij}}{1+\beta_0 z_{ij}}
\end{equation}
(allowing the conditional, exponential model to be recovered in case $\xi_{0}=0$, i.e. when the constraint on the sum of the logarithms of the weights is switched-off) and

\begin{equation}
\langle\ln(w_{ij})|a_{ij}=1\rangle=\psi(1-\xi_0)-\ln(\beta_0+z_{ij}^{-1})
\end{equation}
(see also Appendix A).

\subsubsection{Conditional Pareto model.}

Constraining a slightly more complex function of the weights, i.e. their logarithm, leads to the Hamiltonian

\begin{align}
H(\mathbf{W})&=\sum_{i<j}f(w_{ij}|\xi_{ij})\nonumber\\
&=\sum_{i<j}\xi_{ij}\ln(w_{ij})
\end{align}
which, in turn, induces the distribution

\begin{align}\label{distribution_cpm_2}
Q(\mathbf{W}|\mathbf{A})&=\prod_{i<j}q_{ij}(w_{ij}|a_{ij})\nonumber\\
&=\prod_{i<j}\dfrac{e^{-\xi_{ij}\ln(w_{ij})}}{\zeta_{ij}}\nonumber\\
&=\prod_{i<j}\dfrac{w_{ij}^{-\xi_{ij}}}{\zeta_{ij}}\nonumber\\
&=\prod_{i<j}\dfrac{(\xi_{ij}-1)}{m_{ij}^{1-\xi_{ij}}}w_{ij}^{-\xi_{ij}}
\end{align}
where $m_{ij}$ is the minimum, node pair-specific weight allowed by the model. Each node pair-specific distribution is characterized by a (conditional) expected weight reading

\begin{equation}
\langle w_{ij}|a_{ij}=1\rangle=\left(\dfrac{\xi_{ij}-1}{\xi_{ij}-2}\right)m_{ij}.
\end{equation}

Such a model can be turned into a proper, econometric one by considering the positions

\begin{align}
\xi_{ij}-2&\equiv z_{ij}^{-1},\nonumber\\
m_{ij}&\equiv w_{min}
\end{align}
ensuring that the expected weights are monotonically increasing functions of the gravity specification and leading to the expression

\begin{equation}
\langle w_{ij}|a_{ij}=1\rangle=(1+z_{ij})w_{min}
\end{equation}
where $w_{min}$ is the empirical, minimum weight (see also Appendix A).\\

Let us explicitly notice that the derivation of the gamma and Pareto distributions within the maximum-entropy framework has been already studied in~\cite{Visser2013}; here, however, we aim at making a step further, by individuating a suitable redefinition of these models parameters capable of turning them into proper, econometric ones.

\subsubsection{Conditional log-normal model.}

Adding a global constraint on (a function of) the total variance of the logarithms of the weights to the Hamiltonian defining the Pareto model leads to the expression

\begin{align}
H(\mathbf{W})&=\sum_{i<j}f(w_{ij}|\gamma_0,\xi_{ij})\nonumber\\
&=\sum_{i<j}[\xi_{ij}\ln(w_{ij})+\gamma_0\ln^2(w_{ij})]\nonumber\\
&=\sum_{i<j}\xi_{ij}\ln(w_{ij})+\gamma_0W_3;
\end{align}
the Hamiltonian above induces a distribution reading

\begin{align}\label{distribution_clm}
Q(\mathbf{W}|\mathbf{A})&=\prod_{i<j}q_{ij}(w_{ij}|a_{ij})\nonumber\\
&=\prod_{i<j}\dfrac{e^{-\xi_{ij}\ln(w_{ij})-\gamma_{0}\ln^2(w_{ij})}}{\zeta_{ij}}\nonumber\\
&=\prod_{i<j}\frac{e^{-\xi_{ij}\ln(w_{ij})-\gamma_{0}\ln^2(w_{ij})}}{\sqrt{\frac{\pi}{\gamma_{0}}}e^{\frac{(\xi_{ij}-1)^2}{4\gamma_{0}}}};
\end{align}
each node pair-specific distribution is characterized by a (conditional) expected weight reading

\begin{equation}
\langle w_{ij}|a_{ij}=1\rangle=e^\frac{3-2\xi_{ij}}{4\gamma_0},
\end{equation}
by a (conditional) expected, logarithmic weight reading

\begin{equation}
\langle\ln(w_{ij})|a_{ij}=1\rangle=\frac{1-\xi_{ij}}{2\gamma_0}
\end{equation}
and by a (conditional) logarithmic weight whose squared expectation reads

\begin{equation}
\langle\ln^2(w_{ij})|a_{ij}=1\rangle=\frac{2\gamma_0+(1-\xi_{ij})^2}{4\gamma_0^2}.
\end{equation}

Such a model can be turned into a proper, econometric one by considering the position

\begin{equation}
1-\xi_{ij}\equiv\ln(z_{ij})
\end{equation}
ensuring that the expected weights are monotonically increasing functions of the gravity specification and leading to the expressions

\begin{align}
\langle w_{ij}|a_{ij}=1\rangle&=e^\frac{1+2\ln(z_{ij})}{4\gamma_0},\\
\langle\ln(w_{ij})|a_{ij}=1\rangle&=\frac{\ln(z_{ij})}{2\gamma_{0}},\\
\langle\ln^2(w_{ij})|a_{ij}=1\rangle&=\frac{2\gamma_0+\ln^2(z_{ij})}{4\gamma_0^2}
\end{align}
(see also Appendix A).

\section{Integrated Models\label{sec:integrated}}

MDIP can be also implemented in a straightforward way, by carrying out a constrained optimization of $D_\text{KL}(Q||R)$. In this second case, the following set of constraints

\begin{align}
1&=\int_\mathbb{W}Q(\mathbf{W})d\mathbf{W},\\
\langle C_\alpha\rangle&=\int_\mathbb{W}Q(\mathbf{W})C_\alpha(\mathbf{W})d\mathbf{W},\:\forall\:\alpha
\end{align}
can be specified, with obvious meaning of the symbols. Differentiating the corresponding Lagrangean functional with respect to $Q(\mathbf{W})$ and equating the result to zero leads to

\begin{equation}
Q(\mathbf{W})=\frac{R(\mathbf{W})e^{-H(\mathbf{W})}}{\int_\mathbb{W}R(\mathbf{W})e^{-H(\mathbf{W})}d\mathbf{W}}
\end{equation}
where $H(\mathbf{W})=\sum_\alpha\psi_\alpha C_\alpha$ is, again, the Hamiltonian and $Z=\int_\mathbb{W}e^{-H(\mathbf{W})}d\mathbf{W}$ is the `integrated' partition function.

The explicit functional form of $Q(\mathbf{W})$ can be obtained only once the functional form of both the prior distribution and the constraints has been specified as well. In what follows, we will deal with completely uninformative priors, a choice that amounts at considering the simplified expression

\begin{equation}\label{diff_entropy}
Q(\mathbf{W})=\frac{e^{-H(\mathbf{W})}}{\int_\mathbb{W}e^{-H(\mathbf{W})}d\mathbf{W}};
\end{equation}
notice that the result above could have been also derived by carrying out a constrained minimization of

\begin{equation}
D_\text{KL}(Q||R)=\int_\mathbb{W}Q(\mathbf{W})\ln Q(\mathbf{W})d\mathbf{W}\equiv-S(Q)
\end{equation}
i.e. of (minus) the functional named \emph{differential entropy} into which the KL divergence `degenerates' in case completely uninformative priors are considered.

\subsection{Choosing the constraints}

Let us, now, specify the functional form of the constraints. In what follows, we will deal with a specific instance of the generic Hamiltonian

\begin{align}
H(\mathbf{W})=\sum_{i<j}f(w_{ij}|\alpha_{ij},\beta_{ij});
\end{align}
in particular, one could pose $\alpha_{ij}\equiv \alpha_{0}$, a choice that would lead to constrain the total number of links, or $\alpha_{ij}\equiv\alpha_i+\alpha_j$, a choice that would lead to constrain the whole degree sequence. If not specified otherwise, in what follows we will employ the second functional form and pose $\beta_{ij}\equiv\beta_0+\beta_{ij}$, where $\beta_0$ is the Lagrange multiplier associated with the total weight and $\beta_{ij}$ encodes the dependence on purely econometric quantities. Our choices induce the Hamiltonian of the so-called \emph{integrated exponential model}, i.e.

\begin{align}
H(\mathbf{W})&=\sum_{i<j}f(w_{ij}|\alpha_i+\alpha_j,\beta_0+\beta_{ij})\nonumber\\
&=\sum_{i<j}[(\alpha_i+\alpha_j)a_{ij}+(\beta_0+\beta_{ij})w_{ij}]\nonumber\\
&=\sum_i\alpha_ik_i+\sum_{i<j}(\beta_0+\beta_{ij})w_{ij}\nonumber\\
&=\sum_i\alpha_ik_i+\sum_{i<j}\beta_{ij}w_{ij}+\beta_0W_1\nonumber
\label{hamiltonian_iexp2}
\end{align}
that leads to the distribution

\begin{align}
Q(\mathbf{W})&=\prod_{i<j}q_{ij}(w_{ij})\nonumber\\
&=\prod_{i<j}\dfrac{(x_i x_j)^{a_{ij}}e^{-(\beta_0+\beta_{ij})w_{ij}}}{Z_{ij}}\nonumber\\
&=\prod_{i<j}\dfrac{(x_i x_j)^{a_{ij}}e^{-(\beta_0+\beta_{ij})w_{ij}}}{1+x_i x_j(\beta_0+\beta_{ij})^{-1}}
\end{align}
where $x_i\equiv e^{-\alpha_i}$. The generic node pair-specific distribution induces a probability for nodes $i$ and $j$ to be connected reading

\begin{align}\label{pij_IEXP1_2}
p_{ij}=1-q_{ij}(0)&=\dfrac{x_ix_j(\beta_0+\beta_{ij})^{-1}}{1+x_ix_j(\beta_0+\beta_{ij})^{-1}}\nonumber\\
&=\dfrac{x_ix_j\zeta_{ij}}{1+x_ix_j\zeta_{ij}};
\end{align}
besides, the corresponding expected weight reads

\begin{equation}\label{w_cond_cem42}
\langle w_{ij}\rangle=\frac{p_{ij}}{\beta_0+\beta_{ij}}.
\end{equation}

Equation~\ref{pij_IEXP1_2} clarifies why the models considered in the present section are classified as `integrated': each node pair-specific probability of connection is a function of the parameters controlling for both topological \emph{and} weighted properties. Models of the kind are, thus, capable of `integrating' information concerning a network structure with information concerning its weights, hence employing them in a joint fashion to define both inference steps.

The recipe for the econometric reparametrization of the integrated exponential model can read as the one of its conditional counterpart, i.e.

\begin{equation}
\beta_{ij}\equiv z_{ij}^{-1}
\end{equation}
a position that turns Eq.~\ref{w_cond_cem42} into

\begin{equation}
\langle w_{ij}\rangle=\frac{p_{ij}}{\beta_0+z_{ij}^{-1}}
\end{equation}
where

\begin{equation}
p_{ij}=\frac{x_ix_j}{x_ix_j+\beta_0+z_{ij}^{-1}}
\end{equation}
(see also Appendix B).

\section{Results\label{sec:results}}

The effectiveness of the two classes of models considered in the present paper to reproduce the topological properties of the World Trade Web has been tested on two different datasets, i.e. the Gleditsch one (covering 11 years, from 1990 to 2000~\cite{Gled2002}) and the BACI one (covering 11 years, from 2007 to 2017~\cite{Baci2014}).
To carry out our analyses, we have built the ensemble induced by each model as follows. First, the presence of a link connecting any two nodes $i$ and $j$ is established with probability $p_{ij}$. Numerically, a real number $u$ is drawn from the uniform distribution $U[0,1]$ with unit support and compared with $p_{ij}$: if $u\leq p_{ij}$, then $i$ and $j$ are linked, otherwise they are not. Once the presence of a link is established, it is loaded with a weight by employing the inverse transform sampling technique: another random variable $\eta$, uniformly distributed between 0 and 1, is set equal to the value of the complementary cumulative distribution 
\begin{equation}
F(v_{ij})=\int_0^{v_{ij}}q(w_{ij}|a_{ij}=1)dw_{ij}
\end{equation}
and, inverting the equation $F(v_{ij})=\eta$, one obtains the value of the random variable $v_{ij}$ to be assigned as link weight to the pair $i<j$. Each ensemble is sampled repeatedly to obtain $10^4$ configurations. The error accompanying the estimate of any quantity of interest is quantified via the confidence intervals (CI) induced by the ensemble distribution of the quantity itself.

\begin{table*}[t!]
\begin{subtable}{\textwidth}
\centering
\begin{tabular}{C{2cm}|C{2cm}|C{1cm}|C{1cm}|C{1cm}}
\hline
Dataset & Model & $f^{k_{i}}$ & $f^{k_{nn}}$ & $f^{c_{i}}$ \\
\hline
Gleditsch & I-Exp & $1$ & $1$ & $1$ \\
Gleditsch & UBCM & $1$ & $1$ & $1$ \\
Gleditsch & FM & $1$ & $1$ & $1$ \\
\hline
\end{tabular}
\quad
\begin{tabular}{C{2cm}|C{2cm}|C{1cm}|C{1cm}|C{1cm}}
\hline
Dataset & Model & $f^{k_{i}}$ & $f^{k_{nn}}$ & $f^{c_{i}}$   \\
\hline
BACI & I-Exp & $1$ & $0.09$ & $0.09$ \\
BACI & UBCM & $1$ & $0.09$ & $0.09$ \\
BACI & FM & $0.09$ & $0.09$ & $0.09$ \\
\hline
\end{tabular}
\end{subtable}
\vfill
\begin{subtable}{\textwidth}
\centering
\begin{tabular}{C{2cm}|C{2cm}|C{1cm}|C{1cm}|C{1cm}}
\hline
Dataset & Model & $f^{s_{i}}$ & $f^{s_{nn}}$ & $f^{c_{w}}$ \\
\hline
Gleditsch & I-Exp & $1$ & $1$ & $0.18$ \\
Gleditsch & C-Exp & $1$ & $1$ & $0.18$ \\
Gleditsch & C-Pareto & $0$ & $0$ & $0$ \\
Gleditsch & C-Gamma & $1$ & $1$ & $0.27$ \\
Gleditsch & C-Lognormal & $1$ & $0$ & $0$ \\
\hline
\end{tabular}
\quad
\begin{tabular}{C{2cm}|C{2cm}|C{1cm}|C{1cm}|C{1cm}}
\hline
Dataset & Model & $f^{s_{i}}$ & $f^{s_{nn}}$ & $f^{c_{w}}$ \\
\hline
BACI & I-Exp & $1$ & $1$ & $1$ \\
BACI & C-Exp & $1$ & $1$ & $1$ \\
BACI & C-Pareto & $0$ & $0$ & $0$ \\
BACI & C-Gamma & $1$ & $1$ & $1$ \\
BACI & C-Lognormal & $0$ & $0$ & $0$ \\
\hline
\end{tabular}
\end{subtable}
\caption{Compatibility between the distributions of the expected values of the statistics output by the models considered in the present work and their empirical counterparts for the Gleditsch (left) and the BACI (right) dataset. A large value of $f^s_m$ indicates a large percentage of years for which the distribution of the network statistic predicted by model $m$ is compatible with the empirical one. While all models seem to perform quite well in reproducing the binary statistics on the Gleditsch dataset, this is no longer true when considering the BACI dataset, on which the UBCM outperform the FM - a result that leads us to prefer the former as the `first step-algorithm' of our conditional models. For what concerns the set of weighted statistics, the models constraining $W_1$ (i.e. the integrated exponential model, the conditional exponential model and the conditional gamma model) clearly outperform the others.}
\label{ks_frequency_weighted_table}
\end{table*}

\subsection{Model selection via statistical indicators}

Let us consider two measures of goodness-of-fit, i.e. the reconstruction accuracy $\text{RA}^s_m$ and the Kolmogorov-Smirnov (KS) compatibility frequency $f^s_m$: while $\text{RA}^s_m$ is defined as the percentage of node-specific values of the statistic $s$ falling within the CIs, induced by model $m$, at the significance level of $5\%$, $f^s_m$ measures the percentage of times the distribution of a given, expected statistics $s$, under model $m$, is compatible with the empirical one, according to the two-sample Kolmogorov-Smirnov test; for example, a value $f^s_m=0.8$ would indicate that the distribution of the expected values of the statistics $s$, under model $m$, is found to be compatible with the empirical one on the $80\%$ of the years in the dataset, at the significance level of $5\%$.

\begin{figure*}[t!]
\centering
\begin{subfigure}[t!]{\textwidth}
\centering
\includegraphics[width=\textwidth]{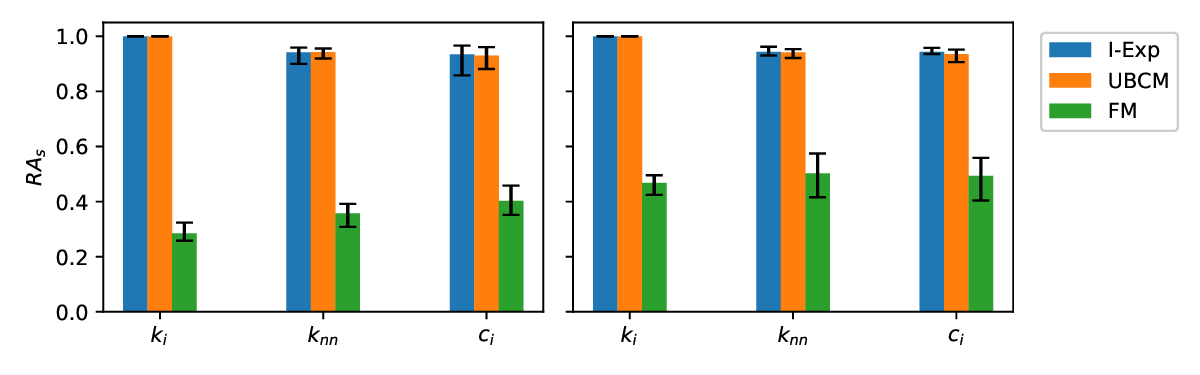}
\caption{\centering Reconstruction accuracy for the binary network statistics.}
\label{fig:binary_ra}
\end{subfigure}
\begin{subfigure}[t!]{\textwidth}
\centering
\includegraphics[width=\textwidth]{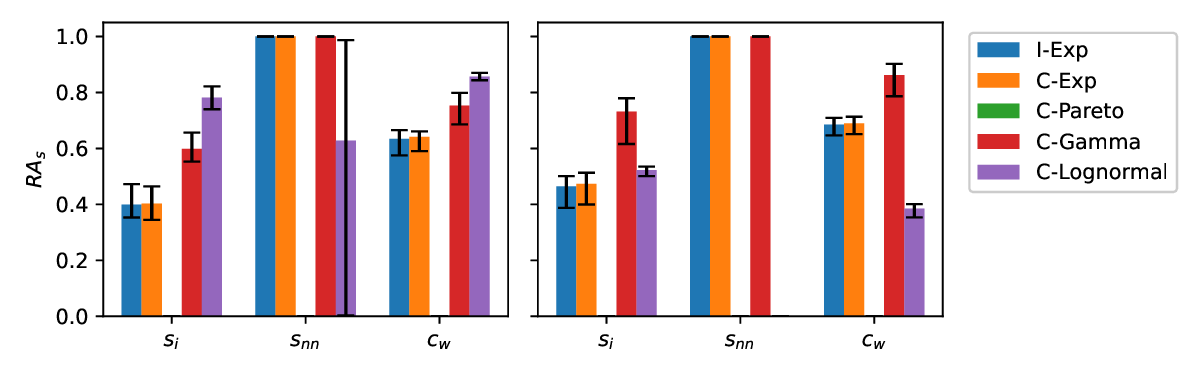}
\caption{\centering Reconstruction accuracy for the weighted network statistics.}
\label{fig:weighted_ra}
\end{subfigure}
\caption{Reconstruction accuracy for the statistics of interest, calculated as the percentage of node-specific values of the statistics $s$ falling within the CI, induced by model $m$, at the significance level of $5\%$; yearly percentages are, then, averaged. The whiskers represent the $2.5$ and $97.5$ percentiles of each $\text{RA}^s_m$ distribution, across different years. Overall, all models perform quite well in reproducing the degrees, with the only exception of the FM - whence our choice of employing the UBCM as the `first step-algorithm' of our conditional models. For what concerns higher-order, binary statistics, both the UBCM and the I-Exp perform quite well while the performance of the FM is much poorer - a result that holds true for both the Gleditsch (left panels) and the BACI (right panels) dataset. For what concerns the weighted statistics, only the average nearest neighbors strength is satisfactorily recovered - however, only by the (integrated and conditional) exponential models and the conditional gamma one.}
\label{fig1}
\end{figure*}

The network statistics for which the values $\text{RA}^s_m$ and $f^s_m$ have been computed are the degree sequence
\begin{equation}
k_i=\sum_{j(\neq i)=1}^Na_{ij},\quad\forall\:i
\end{equation}
(which gives information about the tendency of node $i$ to connect to other trade partners), the average nearest neighbors degree
\begin{equation}
k^{nn}_i=\dfrac{\sum_{j(\neq i)=1}^Na_{ij}k_{j}}{k_{i}},\quad\forall\:i
\end{equation}
(which gives information about the degree correlations), the clustering coefficient
\begin{equation}
c_i=\dfrac{\sum_{j(\neq i)=1}^N\sum_{k(\neq i,j)=1}^Na_{ij}a_{jk}a_{ki}}{k_i(k_i-1)},\quad\forall\:i
\end{equation}
(which counts the percentage of node $i$'s partners that are also partners themselves). For what concerns the weighted statistics, we have considered the strength sequence
\begin{equation}
s_i=\sum_{j(\neq i)=1}^Nw_{ij},\quad\forall\:i
\end{equation}
(which gives information about the trade flow of a country), the average nearest neighbors strength
\begin{equation}
s^{nn}_i=\dfrac{\sum_{j(\neq i)=1}^Na_{ij}s_{j}}{k_i},\quad\forall\:i
\end{equation}
(which gives information about the strength correlations), the weighted clustering coefficient
\begin{equation}
c^w_i=\dfrac{\sum_{j(\neq i)=1}^N\sum_{k(\neq i,j)=1}^Nw_{ij}w_{jk}w_{ki}}{k_i(k_i-1)},\quad\forall\:i
\end{equation}
(that weighs the closed triangular patterns that node $i$ establishes with other trade partners).

\begin{figure*}[t!]
\centering
\includegraphics[width=\textwidth]{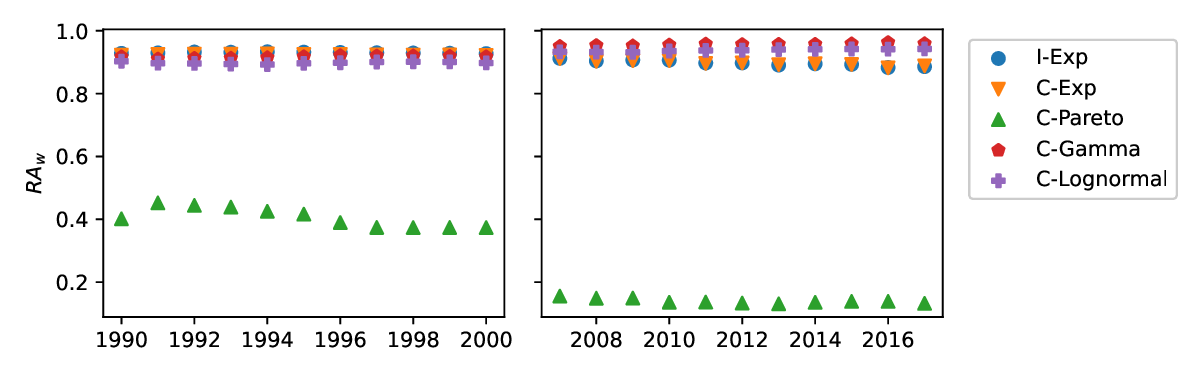}
\caption{Reconstruction accuracy $\text{RA}^w_m$ for the Gleditsch dataset (left panel) and the BACI dataset (right panel). All models perform quite well in reproducing the weights (across all years, on both datasets) with the only exception of the conditional Pareto model. Overall, the best-performing model on the Gleditsch dataset is the integrated exponential while the best-performing model on the BACI dataset is the conditional gamma model.}
\label{fig2}
\end{figure*}

\subsubsection{KS compatibility frequency}

Table~\ref{ks_frequency_weighted_table} lists the values of $f^s_m$ for both binary and weighted network statistics. For what concerns the binary statistics, we report the performance of three different models, i.e. the UBCM, the FM and the integrated exponential model (denoted as I-Exp).

For what concerns the Gleditsch dataset, compatibility is observed for every year; for what concerns the BACI dataset, instead, this is no longer true: in fact, the FM outputs predictions that are not compatible with the empirical values for a large number of years and irrespectively from the considered quantity; the UBCM and the I-Exp (i.e. the models constraining the degrees), instead, output predictions whose compatibility depends on the considered quantity: higher-order statistics are the ones for which the two aforementioned models `fail' to the larger extent. Overall, these results lead us to prefer the UBCM as the `first step-algorithm' of our conditional models.

Let us, now, comment on the performance of our models in reproducing weighted statistics. As it can be appreciated upon looking at Table~\ref{ks_frequency_weighted_table}, the only models outputting predictions whose distributions are compatible with the empirical analogues are the integrated exponential one, the conditional exponential one and the conditional gamma one. On the other hand, employing only logarithmic constraints (as for the conditional Pareto model and the conditional log-normal model) does not help improving the accuracy of the description of the system at hand.

\subsubsection{Reconstruction accuracy}

So far, we have inspected the compatibility of the distributions of the empirical values of each network statistics with the ones of their expected values under each of our models. Let us, now, quantify the extent to which each model is able to recover node-wise information by computing the $\text{RA}^s_m$ values. Figure~\ref{fig1} shows the temporal average of the latter ones (i.e. across the years covered by our datasets), with the whiskers representing their variation, i.e. an indication of the stability of each model performance.

For what concerns the binary statistics (see Fig.~\ref{fig1}a), both the UBCM and the I-Exp perform quite well in reproducing them; on the other hand, the performance of the FM is much poorer. For what concerns the weighted statistics (see Fig.~\ref{fig1}b), only the average nearest neighbors strength is satisfactorily recovered by the (integrated and conditional) exponential models and the conditional gamma one. Still, they are found to perform poorly on the other statistics, i.e. the strength, that is only recovered in distribution on both datasets, and the weighted clustering coefficient, that is only recovered in distribution on the BACI dataset. For what concerns the lognormal model, it performs better than competitors in reproducing the strength and the weighted clustering coefficient on the Gledistch dataset but worse on the BACI dataset, causing its behavior to be dataset-dependent.

Finally, let us inspect the reconstruction accuracy of our models for what concerns our networks link weights. Specifically, let us define $\text{RA}^w_m$, i.e. the percentage of empirical weights falling within the CI, induced by model $m$, at the significance level of $5\%$~\cite{Parisi2020}. Here, we have proceeded numerically, i.e. by considering the $2.5$ and the $97.5$ percentiles induced by the ensemble distribution of each node pair-specific weight. As Fig.~\ref{fig2} shows, all models perform quite well in reproducing the weights (across all years, on both datasets) with the only exception of the conditional Pareto model. Overall, the best-performing model on the Gleditsch dataset is the integrated exponential one while the best-performing model on the BACI dataset is the conditional gamma model.

\subsubsection{Confusion matrix}

The UBCM and the integrated exponential model perform similarly in reproducing the binary statistics, on both datasets. Let us, now, compare them in reproducing the four indicators composing the so-called confusion matrix, i.e. the true positive rate $\langle\text{TPR}\rangle=\langle\text{TP}\rangle/L=\sum_{i<j}a_{ij}p_{ij}/L$ (measuring the percentage of links correctly recovered by a given reconstruction method), the specificity $\langle\text{SPC}\rangle=\langle\text{TN}\rangle/(N(N-1)/2-L)=\sum_{i<j}(1-a_{ij})(1-p_{ij})/(N(N-1)/2-L)$ (measuring the percentage of zeros correctly recovered by a given reconstruction method), the positive predictive value $\langle\text{PPV}\rangle=\langle\text{TP}\rangle/\langle L\rangle=\sum_{i<j}a_{ij}p_{ij}/\langle L\rangle$ (measuring the percentage of links correctly recovered by a given reconstruction method with respect to the total number of links predicted by it) and the accuracy $\langle\text{ACC}\rangle=(\langle\text{TP}\rangle+\langle\text{TN}\rangle)/N(N-1)/2$ (measuring the overall performance of a given reconstruction method in correctly placing both links and zeros).

The results are reported in Table~\ref{tab_accuracy_static}, that shows the increments of the four indicators, defined as $\Delta_X=\langle X\rangle_\text{I-Exp}-\langle X\rangle_\text{UBCM}$ with $X=\text{TPR},\:\text{SPC},\:\text{PPV},\:\text{ACC}$. Notice that each entry of the table is positive, a result signalling that the integrated exponential model steadily performs better than the UBCM. This is further confirmed by the (non-parametric) Wilcoxon rank-sum test on the ensemble distributions of the statistics to compare: all increments are significant, at the $1\%$ level.

\subsection{Model selection via statistical tests}

Let us now rigorously test if constraining the entire degree sequence $\{k_i\}_{i=1}^N$ leads to a significantly better description of our data than that obtainable by just constraining the total number of links $L$.

Upon solving the model constraining the entire degree sequence and the one constraining the total number of links, we are able to construct a vector reading $(X_i^m, Y_i^m)$ where $X_i^m$ is either $\text{RA}^s_m$ or $f^s_m$ for the $i$-th statistics under the `$L$-constrained version' of model $m$; on the other hand, $Y_i^m$ is either $\text{RA}^s_m$ or $f^s_m$ for the $i$-th statistics under the `$k$-constrained version' of model $m$ - naturally, both values have been considered for the same year, keeping the same set of weighted constraints. Pairing statistics as described above allows us to employ the (non-parametric) Wilcoxon signed-rank test for testing the hypotheses $\text{RA}^s_k\leq\text{RA}^s_L$ and $f^s_k\leq f^s_L$, i.e. that the models just constraining $L$ perform better, in reproducing the statistics $s$, than those constraining the entire degree sequence.

Our results let us conclude that, for both datasets, constraining the degree sequence leads to a significant improvement, at the level of $5\%$, of the reconstruction accuracy of the average nearest neighbors degree, the clustering coefficient, the strengths and the average nearest neighbors strength; on the other hand, constraining the degree sequence does not lead to any significant improvement of the reconstruction accuracy of the weighted clustering coefficient. For what concerns the KS compatibility frequency, a significant improvement, at the level of $5\%$, is observed in the description accuracy of the average nearest neighbors degree, the clustering coefficient and the average nearest neighbors strength.

\subsection{Model selection via information criteria}

Let us, now, compare the performance of our models in a more general fashion. To this aim, let us consider the Akaike Information Criterion (AIC) \cite{Akaike1973}, reading

\begin{equation}
\text{AIC}_m=2k-2\mathcal{L}_m
\end{equation}
where $k$ is the number of free parameters of the model and $\mathcal{L}_m$ is its log-likelihood, evaluated at its maximum.

\begin{table*}[t!]
\centering
\begin{tabular}{C{2cm}|C{1cm}|C{1cm}|C{1cm}|C{1cm}}
\hline
Dataset & $\Delta_\text{TPR}$ & $\Delta_\text{SPC}$ & $\Delta_\text{PPV}$ & $\Delta_\text{ACC}$\\
\hline
Gleditsch 90 & $0.017$ & $0.026$ & $0.017$ & $0.020$ \\
Gleditsch 91 & $0.016$ & $0.020$ & $0.016$ & $0.018$ \\
Gleditsch 92 & $0.015$ & $0.019$ & $0.015$ & $0.017$ \\
Gleditsch 93 & $0.015$ & $0.019$ & $0.015$ & $0.017$ \\
Gleditsch 94 & $0.013$ & $0.018$ & $0.013$ & $0.015$ \\
Gleditsch 95 & $0.013$ & $0.019$ & $0.013$ & $0.016$ \\
Gleditsch 96 & $0.012$ & $0.019$ & $0.012$ & $0.015$ \\
Gleditsch 97 & $0.013$ & $0.022$ & $0.013$ & $0.016$ \\
Gleditsch 98 & $0.013$ & $0.022$ & $0.013$ & $0.016$ \\
Gleditsch 99 & $0.013$ & $0.023$ & $0.014$ & $0.017$ \\
Gleditsch 00 & $0.014$ & $0.023$ & $0.014$ & $0.017$ \\
\hline
\end{tabular}
\quad
\begin{tabular}{C{2cm}|C{1cm}|C{1cm}|C{1cm}|C{1cm}}
\hline
Dataset & $\Delta_\text{TPR}$ & $\Delta_\text{SPC}$ & $\Delta_\text{PPV}$ & $\Delta_\text{ACC}$   \\
\hline
BACI 07 & $0.007$ & $0.037$ & $0.007$ & $0.012$ \\
BACI 08 & $0.006$ & $0.035$ & $0.006$ & $0.010$ \\
BACI 09 & $0.005$ & $0.031$ & $0.005$ & $0.009$ \\
BACI 10 & $0.005$ & $0.032$ & $0.005$ & $0.009$ \\
BACI 11 & $0.005$ & $0.029$ & $0.006$ & $0.008$ \\
BACI 12 & $0.005$ & $0.033$ & $0.005$ & $0.009$ \\
BACI 13 & $0.005$ & $0.034$ & $0.005$ & $0.009$ \\
BACI 14 & $0.005$ & $0.031$ & $0.005$ & $0.008$ \\
BACI 15 & $0.005$ & $0.028$ & $0.005$ & $0.008$ \\
BACI 16 & $0.003$ & $0.021$ & $0.004$ & $0.006$ \\
BACI 17 & $0.004$ & $0.028$ & $0.004$ & $0.007$ \\
\hline
\end{tabular}
\caption{Increments of the four indicators composing the confusion matrix, i.e. the true positive rate (TPR), the specificity (SPC), the positive predicted value (PPV) and the accuracy (ACC) when passing from the UBCM to the integrated exponential model for the Gleditsch (left table) and the BACI (right table) datasets. All increments are significant at the $1\%$ level, according to the (non-parametric) Wilcoxon rank-sum test on the ensemble distributions of the statistics to compare.}
\label{tab_accuracy_static}
\end{table*}

\begin{figure*}[t!]
\centering
\begin{subfigure}[t]{\textwidth}
\centering
\includegraphics[width=\textwidth]{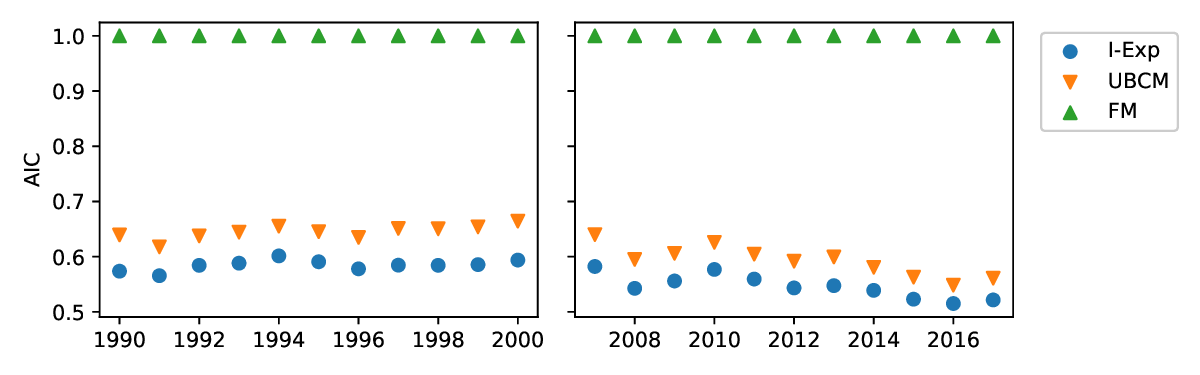}
\caption{\centering AIC values for the binary log-likelihood (normalized by the yearly maximum, across models).}
\label{fig:binary_aic}
\end{subfigure}
\begin{subfigure}[t]{\textwidth}
\centering
\includegraphics[width=\textwidth]{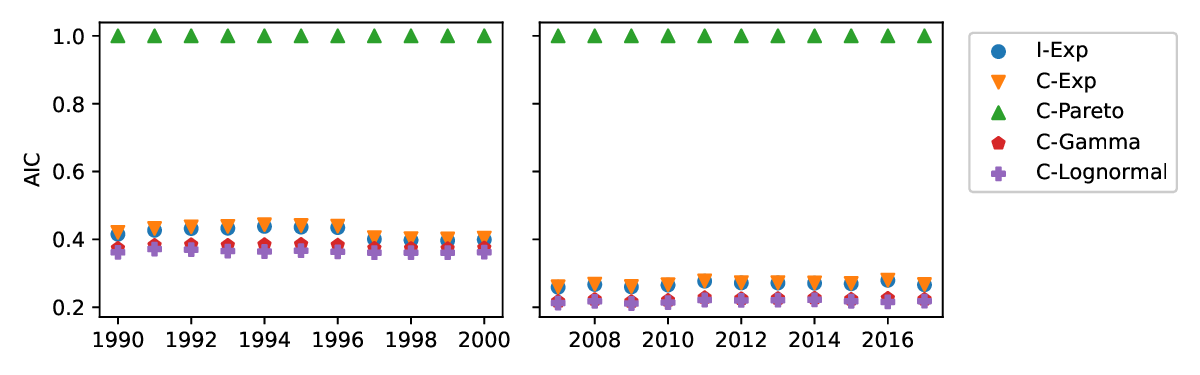}
\caption{\centering AIC values for the full log-likelihood (normalized by the yearly maximum, across models).}
\label{fig:weighted_aic}
\end{subfigure}
\caption{AIC values for the binary and the `full' log-likelihood, normalized by the yearly maximum, across models, for better visualization, for the Gleditsch (left panels) and the BACI (right panel) datasets: a lower AIC value is associated to a better performance. Our plots clearly show that constraining the degree sequence increases a model performance in reproducing a network topology: moreover, the integrated exponential model steadily outperforms the UBCM, signalling that the information gain due to inclusion of economic variables does not affect the parsimony of its description. For what concerns the capability of our models in reproducing weighted properties, the conditional log-normal and gamma models compete, outperforming the other ones.}
\label{fig3}
\end{figure*}

The purely binary log-likelihood induced by model $m$ is readily obtained from Eq.~(\ref{topconditionalL}) and reads

\begin{equation}
\mathcal{L}^{(b)}_m=\ln P(\mathbf{A})=\sum_{i<j}\left[a_{ij}\ln(p_{ij})+(1-a_{ij})\ln(1-p_{ij})\right]
\end{equation}
where $a_{ij}$ is the generic entry of the empirical adjacency matrix and $p_{ij}$ is the model-dependent probability that node $i$ and node $j$ establish a connection. The `binary' AIC values (normalized by the yearly maximum, across models, for better visualization) are reported in Fig.~\ref{fig3}a: the integrated exponential model outperforms the others, across all years, for both datasets. This result suggests that the information gained by including economic factors into the connection probabilities predicted by it does not affect the parsimony of its description, allowing it to perform better than the UBCM.

When, instead, the `full' log-likelihood is considered, reading

\begin{equation}
\mathcal{L}^{(f)}_{\text{I-}m}=\ln Q(\mathbf{W})=\sum_{i<j}\ln(q_{ij}(w_{ij}))
\end{equation}
for integrated models and
 
 \begin{align}
\mathcal{L}^{(f)}_{\text{C-}m}=&\ln P(\mathbf{A})+\ln Q(\mathbf{W}|\mathbf{A})\nonumber\\
=&\sum_{i<j}[a_{ij}\ln(p_{ij})+(1-a_{ij})\ln(1-p_{ij})\nonumber\\
&+\ln(q_{ij}(w_{ij}|a_{ij}))]
\end{align}
for conditional models (see Fig.~\ref{fig3}b), the conditional log-normal and gamma models compete, outperforming the other ones - although the performance of the first one in predicting the network statistics of interest, on the BACI dataset, was less remarkable than that of the competing models (see Fig.~\ref{fig2}b).

\subsection{The Shannon-Fisher plane\label{sec:SF}}

We now complement the analysis of model performance, given in terms of realized likelihood, with an investigation of model `sensitivity', given in terms of the variability of the likelihood across network configurations sampled from the model. To this end, for each conditional model we build the so-called Shannon-Fisher plane~\cite{Vignat2003}, which is a technique that has acquired some popularity in the study of time-series. For instance it has been employed to understand ordinal patterns~\cite{Rosso2012}, quantify the degree of stochasticity~\cite{Ravetti2014}, classify financial stock markets~\cite{Wang2018} and build indicators of economic efficiency~\cite{Fernandes2021}.

Within our context, we can use the Shannon-Fisher technique to project a given model onto a plane by assigning two coordinates to each connected dyad, i.e. to each pair of nodes $(i,j)$ with $a_{ij}=1$, where $a_{ij}$ is taken from the empirical adjacency matrix of the network.
The $y$ coordinate in the plane is the Shannon entropy 
\begin{align}
S_{ij}&=-\int dw\, q_{ij}(w|a_{ij=1})\,\ln q_{ij}(w|a_{ij}=1)\nonumber\\
&=\int dw\,q_{ij}(w|a_{ij}=1)[H_{ij}(w)+\ln\zeta_{ij}]\nonumber\\
&=\langle H_{ij}\rangle+\ln\zeta_{ij},
\end{align}
 which quantifies the degree of uncertainty encoded in the link weight. Note that, since the above entropy is constructed from a continuous pdf, it can attain negative values. This is a well-known problem that can be regularized by introducing the Kullback-Leibler divergence with respect to a continuous uniform pdf, however the result will only consist in an overall shift and rescaling of the $y$ coordinate that are inessential for our discussion below.
 
The $x$ coordinate in the plane is the Fisher Information Measure (FIM), defined as
\begin{align}
F_{ij}&=\int dw\,q_{ij}(w|a_{ij}=1)\left(\frac{\partial\ln\,q_{ij}(w|a_{ij}=1)}{\partial w}\right)^{2}\\
&=\int dw\,q_{ij}(w|a_{ij}=1)\left(\frac{\partial[-H_{ij}(w)-\ln\zeta_{ij}]}{\partial w}\right)^{2}\\
&=\int dw\,q_{ij}(w|a_{ij}=1)\left(\frac{\partial H_{ij}(w)}{\partial w}\right)^{2}\\
&=\int dw\,q_{ij}(w|a_{ij}=1)\left(H_{ij}'(w)\right)^{2}\\
&=\langle(H_{ij}')^{2}\rangle
\end{align}
and quantifying the (average) change in probability induced by small changes in the value of the link weight. 
Notice that the presence of the derivative requires that $q_{ij}(w|a_{ij}=1)$ is continuous throughout the domain of integration, and this is why we consider only conditional models with $a_{ij}=1$ so that there is no `jump' in the unconditional $q_{ij}(w)$ from $w=0$ to $w>0$.
The expression $F_{ij}=\langle(H'_{ij})^2\rangle$ captures the `sensitivity' of the dyadic probability distribution with respect to small changes in the corresponding random variable.
Note that this sensitivity is not captured by Shannon entropy, which is indifferent to any reordering of the values of the random variable, provided each value retains its probability.

\begin{figure*}[t!]
\centering
\includegraphics[width=\textwidth]{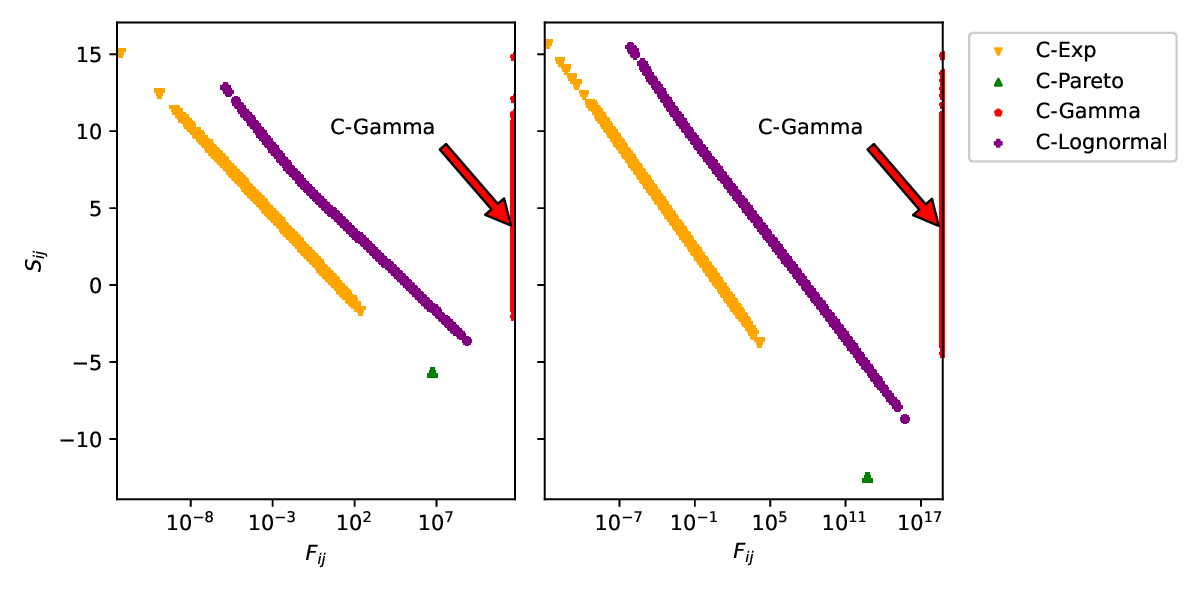}
\caption{Shannon-Fisher plane for each conditional model considered in the present paper. For each pair of nodes $(i,j)$ that are connected in the real network $(a_{ij}=1)$, we consider the conditional weight distribution $q_{ij}(w|a_{ij}=1)$ and plot the corresponding differential Shannon entropy ($y$-axis) versus the Fisher Information Measure ($x$-axis). The results shown correspond to the year 2000 of the Gleditsch dataset (left panel) and to the year 2017 of the BACI dataset (right panel). The dyads of both the conditional exponential and the conditional log-normal model follow a decreasing trend, those of the conditional Pareto model collapse onto a single point, while those of the gamma model are characterized by a diverging Fisher Information Measure independently of their entropy (symbolically depicted as a vertical line at the right edge of the plot). The results show that, while different models (except the Pareto) produce similar values of the entropy, their Fisher measure can be very different (note the logarithmic scale of the $x$-axis). The conditional exponential is the model for which, for a given value of the entropy, the Fisher measure is the minimum one, corresponding to the minimum variability in likelihood across different sampled configurations.}
\label{fig4}
\end{figure*}

In principle, two dyads with the same Shannon entropy can exhibit very different values of the FIM.
This difference is captured by the Shannon-Fisher plane in terms of different positions along the $x$-axis.
In general, since different connected dyads are described by a probability distribution with different parameters, scattering all connected dyads in the plane provides an overall representation of the model identified by $q_{ij}(w|a_{ij}=1)$. Different models are described by different probability distributions and hence have different projections in the Shannon-Fisher plane. In Appendix D we compute the explicit values of $S_{ij}$ and $F_{ij}$ for all the conditional models considered.
Using these calculations we obtain the results shown in Fig.~\ref{fig4} for an illustrative pair of datasets. 
We see that both the conditional exponential model and the conditional log-normal model follow a decreasing pattern. However, the conditional exponential model is, on average, characterized by a smaller FIM, i.e. smaller `sensitivity' to variations of the related random variable.
On the other hand, the conditional Pareto model collapses onto a single point in the Shannon-Fisher plane while the conditional gamma model is characterized by a diverging FIM (because of the divergence of the first two negative moments, see Appendix D).

It is interesting to notice that, if we consider the sum of the $y$ values of all the connected dyads (a sort of `area under the curve') for a given model, we obtain the  Shannon entropy for the entire weighted network, conditional on the empirical binary adjacency matrix $\mathbf{A}$:
\begin{align}
S&=-\int_{\mathbb{W}_\mathbf{A}}Q(\mathbf{W}|\mathbf{A})\ln Q(\mathbf{W}|\mathbf{A})d\mathbf{W}=\sum_{i<j|a_{ij}=1}S_{ij}
\end{align}
(note that the dyadic entropy of $q(w_{ij}|a_{ij}=0)$ is zero, because if $a_{ij}=0$ then $w_{ij}=0$ deterministically).
The above expression also coincides with minus the average likelihood of weighted network  configurations sampled from the model (given the empirical binary structure), hence providing an average (inverse) `goodness of fit' of the weighted model. Similarly, summing the $x$ values of all the connected dyads gives an overall value of the FIM, hence the average change in likelihood of different weighted configurations sampled from the model.
The results shown in Fig.~\ref{fig4} therefore indicate that while different models (except the Pareto) are characterized by similar values of the overall entropy and goodness of fit, the conditional exponential model has minimum overall FIM, thereby producing the most stable outcome (in terms of likelihood of realized configurations) when used to sample weighted networks.

\section{Discussion and conclusions\label{sec:discussion}}

In a companion paper~\cite{Marzio2022} the performance of discrete econometric models in reproducing the structural patterns of the WTW was compared with that of discrete, maximum-entropy ones. The analysis carried out there led to identify the zero-inflated Poisson model as the one performing best among the econometric models; still, it was also found to be largely disfavoured by information criteria such as AIC and BIC. This dilemma has been solved upon looking at a different class of statistical models, i.e. the physics-inspired ones: the latter have been found to outperform the purely econometric ones for reconstruction purposes, the reason lying in the higher accuracy achieved by them in estimating the topological structure of networks.

With this contribution, we extend the work carried out in~\cite{Marzio2022} by, first, introducing models to infer the topology and the weights of (undirected, weighted) networks defined by continuous-valued data and, then, turning them into proper, econometric ones. In order to do so, we present a theoretical, physics-inspired framework based upon the constrained minimization of the KL divergence - hence, implementing the Minimum Discrimination Information Principle, that generalizes the Maximum-Entropy Principle - and capable of accommodating both integrated and conditional (continuous) models.

The main difference between the models belonging to these classes lies in the way the estimation of the topology is carried out; while conditional models disentangle the purely binary step from the (conditional) weighted one, integrated models do not, letting both topological and weighted constraints determine all relevant, structural features of a network. An example of integrated model is provided by the Enhanced Configuration Model (ECM), defined by constraints such as the degree and the strength sequences and described by a mixed Bernoulli-geometric~\cite{Mastrandrea2014a,Mastrandrea2014b} (also called Bose-Fermi~\cite{Garlaschelli2009}) distribution; examples of continuous, conditional models are provided by the $\text{CReM}_\text{A}$ and the $\text{CReM}_\text{B}$~\cite{Parisi2020}. From a more econometric perspective, hurdle models are conditional in nature while zero-inflated models can be thought as integrated, the estimation steps being carried out by selecting a distribution out of a basket of available ones.

Our analysis leads to several conclusions: 1) constraining the entire degree sequence leads to a statistically significant improvement in the reconstruction accuracy of the WTW. In particular, the integrated exponential model, described by the Hamiltonian $H(\mathbf{W})=\sum_i\alpha_ik_i+\sum_{i<j}\beta_{ij}w_{ij}+\beta_0 W_1$, provides a very accurate, structural reconstruction while being favoured by information criteria: although it is defined by $N+1$ purely topological constraints, AIC reveals them as `irreducible', i.e. necessary to provide a satisfactory explanation of the network generating process; 2) when considering weighted quantities, the conditional gamma model is the one performing best (although it competes with the integrated exponential one in reproducing properties such as the weights, on some of the temporal snapshots covered by our datasets), according to information criteria. To be noticed, however, that if strengths are not explicitly constrained - jointly with the degrees - maximum-entropy models recover them only `in distribution' while failing to reproduce their exact values. The same consideration holds true for the weighted clustering coefficient.

Coming to comparing the models belonging to the classes considered in the present work, the two, best-performing ones are the integrated exponential model and the conditional gamma model, i.e. the ones constraining the total weight (although the conditional exponential model constrains the total weight as well, it is outperformed by the conditional gamma one within the class of conditional model): hence, $W_1$ seems to constitute a somehow fundamental quantity to be necessarily accounted for in order to achieve a good reconstruction accuracy. From an economic point of view, the parameter $\beta_0$ constraining the total weight can be interpreted as a sort of `shadow price' to be paid by everyone to exchange goods.

Additional information is provided by our analysis of the Shannon-Fisher plane, which combines Shannon entropy, i.e. the (inverse)  likelihood of a model, with the Fisher Information Measure, i.e. the average variability of the likelihood itself across different sampled configurations.
The conditional exponential model turns out the be the least variable in likelihood, hence the most stable.
It is worth noticing at this point that our maximum-entropy approach is formulated for canonical ensembles, i.e. for `soft constraints', which implies that different realizations of the network have fluctuating values of the weighted sufficient statistics. These fluctuations are the origin of the FIM.
By contrast, if we were to formulate microcanonical models with `hard constraints', then the sufficient statistics would not fluctuate and the overall FIM would be zero.
Therefore the Shannon-Fisher plane shows that, among the canonical models considered here, the conditional exponential is the closest to the `least soft' extreme, while the conditional gamma is at the opposite `softest' extreme where the FIM diverges. 
As a question left for future research, it would be interesting to relate the behaviour of the FIM to the phenomenon of inequivalence of canonical and microcanonical ensembles of networks~\cite{squartini2015breaking}.

Overall, we believe the framework proposed in this contribution to have the potential of reconciling the approach adopted by network scientists for reconstructing economic networks, and focusing on the purely structural aspects of a network formation, with the approach characterizing econometrics, tailored to inform these same models with macro-economic quantities - in all cases considered here, purely bilateral ones such as the GDPs and the geographic distances. From an operative point of view, our (classes of) models combine the pros of both approaches: the importance of purely structural information (highlighted by physics-inspired models) can be accounted for by constraining the entire degree sequence; on top of that, a second step is needed to estimate a network weighted structure. Although the information provided by the total weight cannot be discarded without affecting the overall performance of a model, such an estimation can rests upon econometric considerations driving the reparametrization of otherwise purely structural models.

\section{The DyGyS Python package}

As an additional result, we release a Python package named `DyGyS - DYadic GravitY regression models with Soft constraints' and containing routines to implement all models considered in the present work as well as those considered in the companion paper~\cite{Marzio2022}. The package is available at the following URL: \href{https://github.com/MarsMDK/DyGyS}{\texttt{https://github.com/MarsMDK/DyGyS}}.

\section*{Acknowledgements}

This work is supported by the European Union – Horizon 2020 Program under the scheme “INFRAIA-01-2018-2019 – Integrating Activities for Advanced Communities”, Grant Agreement n.871042, “SoBigData++: European Integrated Infrastructure for Social Mining and Big Data Analytics”(\href{http://www.sobigdata.eu}{http://www.sobigdata.eu}). This work has been also supported by the project `Network analysis of economic and financial resilience', Italian DM n. 289, 25-03-2021 (PRO3 Scuole) CUP D67G22000130001. DG acknowledges support from the Dutch Econophysics Foundation (Stichting Econophysics, Leiden, the Netherlands) and the Netherlands Organization for Scientific Research (NWO/OCW). MDV and DG acknowledge support from the `Programma di Attivit\`a Integrata' (PAI) project `Prosociality, Cognition and Peer Effects' (Pro.Co.P.E.), funded by IMT School for Advanced Studies.

\section*{Appendix A - Conditional models}

Any member of the class of conditional models is described by the expression 

\begin{equation}
Q(\mathbf{W}|\mathbf{A})=\dfrac{e^{-H(\mathbf{W})}}{\int_{\mathbb{W}_\mathbb{A}}e^{-H(\mathbf{W})}d\mathbf{W}} 
\end{equation}
the single instances being characterized by different expressions of the Hamiltonian. Since, however, each Hamiltonian considered here is a sum over the node pairs, the result

\begin{align}
Q(\mathbf{W}|\mathbf{A})&=\dfrac{e^{-\sum_{i<j}H_{ij}(w_{ij})}}{\int_{\mathbb{W}_\mathbb{A}}e^{-\sum_{i<j}H_{ij}(w_{ij})}d\mathbf{W}}\nonumber\\
&=\prod_{i<j}\dfrac{e^{-H_{ij}(w_{ij})}}{\left[\int_{m_{ij}}^{+\infty}e^{-H_{ij}(w_{ij})}dw_{ij}\right]^{a_{ij}}}\nonumber\\
&=\prod_{i<j}\dfrac{e^{-H_{ij}(w_{ij})}}{\zeta_{ij}^{a_{ij}}}
\end{align}
holds irrespectively from the specific functional form of $H_{ij}(w_{ij})$. Notice that $m_{ij}$ is the minimum, pair-specific weight allowed by the model. The identifications

\begin{align}
H_{ij}(w_{ij})\equiv f(w_{ij}|\underline{\theta},z_{ij}(\underline{\psi}))
\end{align}
and

\begin{equation}
\mathcal{L}\equiv\ln Q(\mathbf{W}|\mathbf{A})
\end{equation}
lead to estimate parameters by solving the (coupled) systems of purely structural equations $\frac{\partial\mathcal{L}}{\partial\underline{\theta}}=\underline{0}$ and econometric-like equations $\frac{\partial\mathcal{L}}{\partial\underline{\tau}}=\underline{0}$, i.e.

\begin{equation}
\begin{cases}
\sum_{i<j|a_{ij}=1}\left[\frac{\partial f(w_{ij}|\underline{\theta},z_{ij}(\underline{\psi}))}{\partial\underline{\theta}}+\frac{1}{\zeta_{ij}}\left(\frac{\partial\zeta_{ij}}{\partial\underline{\theta}}\right)\right]=\underline{0},\\
\sum_{i<j|a_{ij}=1}\left[\frac{\partial f(w_{ij}|\underline{\theta},z_{ij}(\underline{\psi}))}{\partial z_{ij}}+\frac{1}{\zeta_{ij}}\left(\frac{\partial\zeta_{ij}}{\partial z_{ij}}\right)\right]\frac{\partial z_{ij}}{\partial\underline{\psi}}=\underline{0}.
\end{cases}
\end{equation}

Notice that the estimation of the parameters carried out by maximizing the conditional likelihood, and letting only the positive weights to be accounted for, is perfectly consistent with the theory of hurdle models~\cite{Cragg1971,Mullahy1986}: although alternative estimation procedures can be devised (see, for example,~\cite{Parisi2020}), in the present paper, we will stick to the proper, econometric one - which has been already employed in our companion paper~\cite{Marzio2022}, to estimate the parameters of conditional, discrete-valued models.

\subsection{Conditional exponential model}

The conditional exponential model is defined by the expression

\begin{equation}
H_{ij}(w_{ij})=(\beta_0+\beta_{ij})w_{ij}
\end{equation}
that induces the following node pair-specific partition function

\begin{equation}
\zeta_{ij}=\int_0^{+\infty}e^{-(\beta_0+\beta_{ij})w_{ij}}dw_{ij}=\frac{1}{\beta_0+\beta_{ij}}.
\end{equation}

After the econometric reparametrization, according to which $\beta_{ij}\equiv z_{ij}^{-1}$, the log-likelihood function of the conditional exponential model reads

\begin{equation}
\mathcal{L}=\sum_{\substack{i<j\\(a_{ij}=1)}}[-(\beta_0+z_{ij}^{-1})w_{ij}-\ln(\zeta_{ij})];
\end{equation}
hence, its maximization leads to the system of equations

\begin{equation}
\begin{cases}
\sum_{i<j|a_{ij}=1}[\langle w_{ij}|a_{ij}=1\rangle-w_{ij}]&=0\\
\sum_{i<j|a_{ij}=1}[\langle w_{ij}|a_{ij}=1\rangle-w_{ij}]\partial_{\underline{\alpha}}(z_{ij}^{-1})&=0
\end{cases}
\end{equation}
where $\langle w_{ij}|a_{ij}=1\rangle=\frac{z_{ij}}{1+\beta_0 z_{ij}}$. Notice that we have a condition on the parameters, reading $\beta_0+\beta_{ij}>0$.

\subsection{Conditional gamma model}

The conditional gamma model is defined by the expression

\begin{equation}
H_{ij}(w_{ij})=(\beta_0+\beta_{ij})w_{ij}+\xi_0\ln(w_{ij})
\end{equation}
and induces the following node pair-specific partition function

\begin{align}
\zeta_{ij}&=\int_0^{\infty}e^{-(\beta_0+\beta_{ij})w_{ij}}w_{ij}^{-\xi_{0}}dw_{ij}\nonumber\\
&=\frac{\Gamma(1-\xi_0)}{(\beta_0+\beta_{ij})^{1-\xi_0}}.
\end{align}

After the econometric reparametrization, according to which $\beta_{ij}\equiv z_{ij}^{-1}$, the log-likelihood function of the conditional gamma model reads

\begin{align}
\mathcal{L}&=\sum_{\substack{i<j\\(a_{ij}=1)}}[-(\beta_0+z_{ij}^{-1})w_{ij}-\xi_0\ln(w_{ij})-\ln(\zeta_{ij})];
\end{align}
hence, its maximization leads to the system of equations

\begin{equation}
\begin{cases}
\sum_{i<j|a_{ij}=1}[\langle w_{ij}|a_{ij}=1\rangle-w_{ij}]&=0\\
\sum_{i<j|a_{ij}=1}[\langle\ln(w_{ij})|a_{ij}=1\rangle-\ln(w_{ij})]&=0\\
\sum_{i<j|a_{ij}=1}[\langle w_{ij}|a_{ij}=1\rangle-w_{ij}]\partial_{\underline{\alpha}}(z_{ij}^{-1})&=0
\end{cases}
\end{equation}
where $\langle w_{ij}|a_{ij}=1\rangle=\frac{z_{ij}(1-\xi_0)}{1+\beta_0 z_{ij}}$ and $\langle\ln(w_{ij})|a_{ij}=1\rangle=\psi(1-\xi_0)-\ln(\beta_0+z_{ij}^{-1})$. Notice that we have conditions on the parameters, reading $\beta_0+\beta_{ij}>0$ and $\xi_0<1$.

\subsection{Conditional Pareto model}

The conditional Pareto model is defined by the expression

\begin{equation}
H_{ij}(w_{ij})=\xi_{ij}\ln(w_{ij})
\end{equation}
that induces the following node pair-specific partition function

\begin{align}
\zeta_{ij}&=\int_{m_{ij}}^{+\infty} e^{-\xi_{ij}\ln(w_{ij})}dw_{ij}\nonumber\\
&=\int_{m_{ij}}^{+\infty}w_{ij}^{-\xi_{ij}}dw_{ij}\nonumber\\
&=\frac{m_{ij}^{1-\xi_{ij}}}{\xi_{ij}-1}.
\end{align}

After the econometric reparametrization, according to which $\xi_{ij}-2 \equiv z_{ij}^{-1}$ and $m_{ij} \equiv w_{min}$, the log-likelihood function of the conditional Pareto model reads

\begin{align}
\mathcal{L}=\sum_{\substack{i<j\\(a_{ij}=1)}}[-(2+z_{ij}^{-1})\ln(w_{ij})-\ln(\zeta_{ij})];
\end{align}
hence, its maximization leads to the system of equations

\begin{equation}
\sum_{i<j|a_{ij}=1}[\langle\ln(w_{ij})|a_{ij}=1\rangle-\ln(w_{ij})]\partial_{\underline{\alpha}}(z_{ij}^{-1})=0
\end{equation}
where $\langle \ln(w_{ij})|a_{ij}=1\rangle=\ln(w_{min})+\frac{z_{ij}}{1+z_{ij}}$. Notice that we have conditions on the parameters, reading $w_{min}>0$ and $\xi_{ij}>2$.

\subsection{Conditional log-normal model}

The conditional log-normal model is defined by the expression

\begin{align}
H_{ij}(w_{ij})=\xi_{ij}\ln(w_{ij})+\gamma_{0}\ln^2(w_{ij})
\end{align}
that induces the following node pair-specific partition function

\begin{align}
\zeta_{ij}&=\int_0^{+\infty}e^{-\xi_{ij}\ln(w_{ij})-\gamma_{0}\ln^2(w_{ij})}dw_{ij}\nonumber\\
&=\int_{-\infty}^{+\infty}e^{(1-\xi_{ij})t_{ij}}e^{-\gamma_{0}t_{ij}^2}dt_{ij}\nonumber\\
&=\sqrt{\frac{\pi}{\gamma_{0}}}e^{\frac{(\xi_{ij}-1)^2}{4\gamma_{0}}}
\end{align}
a result that is readily obtained by putting $t_{ij}=\ln(w_{ij})$ and exploiting the relationship $\int_{-\infty}^{+\infty}e^{-ax^2+bx+c}dx=\sqrt{\frac{\pi}{a}}e^{\frac{b^2}{4a}+c}$.

After the econometric reparametrization, according to which $1-\xi_{ij}\equiv\ln(z_{ij})$, the log-likelihood function of the conditional log-normal model reads

\begin{align}
\mathcal{L}=\sum_{\substack{i<j\\(a_{ij}=1)}}[(\ln(z_{ij})+1)\ln(w_{ij})-\gamma_{0} \ln^2(w_{ij})-\ln(\zeta_{ij})];
\end{align}
hence, its maximization leads to the system of equations

\begin{equation}
\begin{cases}
\sum_{i<j|a_{ij}=1}[\langle\ln^2(w_{ij})|a_{ij}=1\rangle-\ln^2(w_{ij})]&=0\\
\sum_{i<j|a_{ij}=1}[\langle\ln(w_{ij})|a_{ij}=1\rangle-\ln(w_{ij})]\partial_{\underline{\alpha}}\ln(z_{ij})&=0
\end{cases}
\end{equation}
where $\langle\ln^2(w_{ij})|a_{ij}=1\rangle =\frac{2\gamma_0+\ln^2(z_{ij})}{4\gamma_0^2}$ and $\langle\ln(w_{ij})|a_{ij}=1\rangle=\frac{\ln(z_{ij})}{2\gamma_{0}}$. Notice that we have a condition on the parameters, reading $\gamma_0>0$.

\section*{Appendix B - Integrated models}

Any member of the class of integrated models is described by the expression

\begin{equation}
Q(\mathbf{W})=\dfrac{e^{-H(\mathbf{W})}}{\int_{\mathbb{W}}e^{-H(\mathbf{W})}d\mathbf{W}} 
\end{equation}
the single instances being characterized by different expressions of the Hamiltonian. Since, however, each Hamiltonian considered here is a sum over the node pairs, the result

\begin{align}
Q(\mathbf{W})&=\dfrac{e^{-\sum_{i<j}H_{ij}(w_{ij})}}{\int_{\mathbb{W}}e^{-\sum_{i,j}H_{ij}(w_{ij})}d\mathbf{W}}\nonumber\\
&=\prod_{i<j}\dfrac{e^{-H_{ij}(w_{ij})}}{\sum_{a_{ij}=0,1}\int_{\Theta[w_{ij}]=a_{ij}}e^{-H_{ij}(w_{ij})}dw_{ij}}\nonumber\\
&=\prod_{i<j}\dfrac{e^{-H_{ij}(w_{ij})}}{Z_{ij}}
\end{align}
holds irrespectively from the specific functional form of $H_{ij}(w_{ij})$. The identifications

\begin{align}
H_{ij}(w_{ij})\equiv f(w_{ij}|\underline{\theta},z_{ij}(\underline{\psi}))
\end{align}
and

\begin{equation}
\mathcal{L}\equiv\ln Q(\mathbf{W})
\end{equation}
lead to estimate parameters by solving the (coupled) systems of purely structural equations $\frac{\partial\mathcal{L}}{\partial\underline{\theta}}=\underline{0}$ and econometric-like equations $\frac{\partial\mathcal{L}}{\partial\underline{\psi}}=\underline{0}$, i.e.

\begin{equation}
\begin{cases}
\sum_{i<j}\left[\frac{\partial f(w_{ij}|\underline{\theta},z_{ij}(\underline{\psi}))}{\partial\underline{\theta}}+\frac{1}{Z_{ij}}\left(\frac{\partial Z_{ij}}{\partial\underline{\theta}}\right)\right]=\underline{0},\\
\sum_{i<j}\left[\frac{\partial f(w_{ij}|\underline{\theta},z_{ij}(\underline{\psi}))}{\partial z_{ij}}+\frac{1}{Z_{ij}}\left(\frac{\partial Z_{ij}}{\partial z_{ij}}\right)\right]\frac{\partial z_{ij}}{\partial\underline{\psi}}=\underline{0}.
\end{cases}
\end{equation}

The integrated exponential model we have considered in the present paper is defined by the expression

\begin{equation}
H_{ij}(w_{ij})=(\alpha_i+\alpha_j)a_{ij}+(\beta_0+\beta_{ij})w_{ij}
\end{equation}
that induces the following node pair-specific partition function

\begin{align}
Z_{ij}&=\sum_{a_{ij}=0}^1\int_{\Theta[w_{ij}]=a_{ij}}e^{-(\alpha_i+\alpha_j)a_{ij}-(\beta_0+\beta_{ij})w_{ij}}dw_{ij}\nonumber\\
&=1+e^{-(\alpha_i+\alpha_j)}\int_0^{+\infty}e^{-(\beta_0+\beta_{ij})w_{ij}}dw_{ij}\nonumber\\
&=1+\frac{e^{-(\alpha_i+\alpha_j)}}{\beta_0+\beta_{ij}}.
\end{align}

After the econometric reparametrization, according to which $\beta_{ij}\equiv z_{ij}^{-1}$, the log-likelihood function of the exponential model reads

\begin{equation}
\mathcal{L}=\sum_{i<j}[-(\alpha_i+\alpha_j)a_{ij}-(\beta_0+z_{ij}^{-1})w_{ij}-\ln(Z_{ij})];
\end{equation}
hence, its maximization leads to the system of equations

\begin{equation}
\begin{cases}
\langle k_i\rangle-k_{i}&=0,\:\forall\:i\\
\langle W\rangle-W&=0\\
\sum_{i<j}[\langle w_{ij}\rangle-w_{ij}]\partial_{\underline{\alpha}}(z_{ij}^{-1})&=0\\
\end{cases}
\end{equation}
where $k_i=\sum_{j(\neq i)}a_{ij}$ is the empirical degree of node $i$, $w_{ij}$ is the empirical, pair-specific weight and $W=\sum_{i<j}w_{ij}$ is the empirical, total weight; $\langle k_i\rangle=\sum_{j(\neq i)}p_{ij}$, $\langle w_{ij}\rangle$ and $\langle W\rangle=\sum_{i<j}\langle w_{ij}\rangle$ are their expected counterparts.

Notice that we have a condition on the parameters, reading $\beta_0+\beta_{ij}>0$.

\section*{Appendix C - Turning structural models into econometric models}

So far, we have derived two classes of models, by explicitly solving the constrained maximization of a number of functionals derived from the KL divergence. As the functional form of the probability distributions belonging to the two classes (solely) depends on the enforced constraints, such models `are born' as purely structural ones.

In order to turn them into candidate models to be employed for econometric purposes, we need to properly transform (some of) the Lagrange multipliers into functions of the econometric quantities of relevance for the problem at hand. In this respect, the theory of GLMs provides helpful suggestions about how to proceed; besides, one can figure out some (sets of) basic requirements such a transformation should satisfy:

\begin{itemize}
\item the transformation should turn the expected values $\langle w_{ij}\rangle$ and $\langle w_{ij}|a_{ij}=1\rangle$ into positive, monotonically increasing functions of $z_{ij}$;

\item the transformation should not violate the mathematical requirements to have well-defined (first and second) distribution moments.
\end{itemize}

In what follows, we will focus on the conditional models.\\

\noindent\emph{Conditional exponential model.} It is characterized by the expression

\begin{equation}
\langle w_{ij}|a_{ij}=1\rangle=\frac{1}{\beta_0+\beta_{ij}}
\end{equation}
that can be turned into an econometric one by posing 

\begin{equation}
\langle w_{ij}|a_{ij}=1\rangle=\frac{1}{\beta_0+\beta_{ij}}\equiv
g(z_{ij})
\end{equation}
according to the prescription informing the so-called generalized linear models (GLMs). Before specifying the functional form of $g$, let us consider that the dyadic parameter $\beta_{ij}$ must be decreasing in $z_{ij}$ - a requirement that can be justified upon identifying $\beta_{ij}$ as the `shadow price' that countries $i$ and $j$ have to pay to trade a unity of goods \cite{Bargigli2016}; analogously, $\beta_0$ can be interpreted as modelling a global tax that everyone has to pay to exchange goods - independently of its trade `capacity'. These considerations lead us to impose $\beta_{ij}\equiv z_{ij}^{-1}$, a choice inducing the expression

\begin{equation}
\langle w_{ij}|a_{ij}=1\rangle=\frac{1}{\beta_0+z_{ij}^{-1}}=\frac{z_{ij}}{1+\beta_0 z_{ij}}
\end{equation}
which violates none of the requirements listed at the beginning of the section.\\

\noindent\emph{Conditional gamma model.} It is characterized by the expressions
\begin{align}
\langle w_{ij}|a_{ij}=1\rangle&=\frac{1-\xi_0}{\beta_0+\beta_{ij}},\\
\langle\ln(w_{ij})|a_{ij}=1\rangle&=\psi(1-\xi_0)-\ln(\beta_0+\beta_{ij})
\end{align}
(where $\psi(x)=\Gamma'(x)/\Gamma(x)$ is the digamma function) that can be turned into econometric ones by posing $\beta_{ij}\equiv z_{ij}^{-1}$, according to considerations which are analogous to those driving the econometric reparametrization of the conditional, exponential model. This choice induces the expressions

\begin{align}
\langle w_{ij}|a_{ij}=1\rangle&=\frac{1-\xi_0}{\beta_0+z_{ij}^{-1}}=\frac{(1-\xi_0)z_{ij}}{1+\beta_0 z_{ij}},\\
\langle\ln(w_{ij})|a_{ij}=1\rangle&=\psi(1-\xi_0)-\ln(\beta_0+z_{ij}^{-1});
\end{align}
notice that the conditional, exponential model is recovered in case $\xi_{0}=0$ (i.e. when the constraint on the sum of the logarithms of weights is switched-off).\\

\noindent\emph{Conditional Pareto model.} It is characterized by the expression

\begin{equation}
\langle w_{ij}|a_{ij}=1\rangle=\left(\dfrac{\xi_{ij}-1}{\xi_{ij}-2}\right)m_{ij}
\end{equation}
that can be turned into an econometric one by posing

\begin{equation}
\langle w_{ij}|a_{ij}=1\rangle=\left(\dfrac{\xi_{ij}-1}{\xi_{ij}-2}\right)m_{ij}\equiv g(z_{ij})
\end{equation}
according to the prescription informing the GLMs. Upon considering that 1) the (conditional) expected value is well defined only if $\xi_{ij}-2>0$ and that 2) a linear relationship between the former and $z_{ij}$ would be desirable, a suitable reparametrization may read $\xi_{ij}-2\equiv z_{ij}^{-1}$ and $m_{ij}\equiv w_{min}$, in turn leading to

\begin{equation}
\langle w_{ij}|a_{ij}=1\rangle=(1+z_{ij})w_{min}
\end{equation}
which violates none of the requirements listed at the beginning of the section.\\

\noindent\emph{Conditional log-normal model.} It is characterized by the expressions

\begin{align}
\langle\ln(w_{ij})|a_{ij}=1\rangle&=\frac{1-\xi_{ij}}{2\gamma_0},\\
\langle\ln^2(w_{ij})|a_{ij}=1\rangle&=\frac{2\gamma_0+(1-\xi_{ij})^2}{4\gamma_0^2}.
\end{align}

Upon considering that the logarithm of weights can admit negative values, i.e. when $w_{ij}\in(0,1)$, and that there are no theoretical restrictions on the sign of $\langle\ln(w_{ij})|a_{ij}=1\rangle$, a suitable reparametrization may read $1-\xi_{ij}\equiv\ln(z_{ij})$, in turn leading to

\begin{align}
\langle\ln(w_{ij})|a_{ij}=1\rangle&=\frac{\ln(z_{ij})}{2\gamma_0},\\
\langle\ln^2(w_{ij})|a_{ij}=1\rangle&=\frac{2\gamma_0+\ln^2(z_{ij})}{4\gamma_0^2}
\end{align}
which violate none of the requirements listed at the beginning of the section.

\section*{Appendix D - The Shannon-Fisher plane}

Here, starting from the conditional probability density function $q_{ij}(w|a_{ij}=1)$, for each connected dyad we compute explicitly the continuous Shannon entropy $S_{ij}$ and the Fisher Information Measure (FIM) $F_{ij}$ needed to construct the Shannon-Fisher plane introduced in Sec.~\ref{sec:SF}. We do so for each model separately.

\subsubsection{Conditional exponential model}

The conditional exponential model is defined by the probability distribution

\begin{equation}
q_{ij}(w_{ij}|a_{ij}=1)=(\beta_0+\beta_{ij})e^{-(\beta_0+\beta_{ij})w_{ij}}
\end{equation}
inducing a Shannon entropy reading

\begin{equation}
S_{ij}=\langle H_{ij}\rangle+\ln\zeta_{ij}=1-\ln[\beta_0+\beta_{ij}]
\end{equation}
and a FIM reading

\begin{equation}
F_{ij}=\langle(H'_{ij})^2\rangle=(\beta_{0}+\beta_{ij})^2;
\end{equation}
as the value of the parameter $\beta_{0}+\beta_{ij}$ increases, Shannon entropy decreases while Fisher Information Measure increases as well.

\subsubsection{Conditional gamma model}

The conditional gamma model is defined by the probability distribution

\begin{equation}
q_{ij}(w_{ij}|a_{ij}=1)=\frac{(\beta_0 + \beta_{ij})^{1-\xi_0}}{\Gamma(1-\xi_0)}w_{ij}^{-\xi_0}e^{-(\beta_0+\beta_{ij})w_{ij}}
\end{equation}
inducing a Shannon entropy reading

\begin{align}
S_{ij}&=\langle H_{ij}\rangle+\ln\zeta_{ij}\nonumber\\
&=-\ln[\beta_{0}+\beta_{ij}]+\xi_{0}\psi(1-\xi_{0})+\ln\Gamma(1-\xi_{0})+(1-\xi_{0})
\end{align}
and a FIM reading

\begin{align}
F_{ij}&=\langle(H'_{ij})^2\rangle=(\beta_{0}+\beta_{ij})^2\nonumber\\
&+2\xi_{0}(\beta_{0}+\beta_{ij})\langle w_{ij}^{-1}|a_{ij}=1\rangle+\xi_{0}^{2}\langle w_{ij}^{-2}|a_{ij}=1\rangle\nonumber\\
&=(\beta_{0}+\beta_{ij})^2\left[1+2\xi_{0}\dfrac{\Gamma(-\xi_{0})}{\Gamma(1-\xi_{0})}+\xi_{0}^2\frac{\Gamma(-1-\xi_{0})}{\Gamma(1-\xi_{0})}\right];
\end{align}
the expression above does not diverge for the values of the parameter $\xi_0$ ensuring that the (first) two, negative moments, $\langle w_{ij}^{-1}|a_{ij}=1\rangle$ and $\langle w_{ij}^{-2}|a_{ij}=1\rangle$, of the (conditional) gamma distribution do not diverge as well, i.e. $\xi_{0}<-1$.

\subsubsection{Conditional Pareto model}

The conditional Pareto model is defined by the probability distribution

\begin{equation}
q_{ij}(w_{ij}|a_{ij}=1)=\frac{\xi_{ij}-1}{m_{ij}^{1-\xi_{ij}}} w_{ij}^{-\xi_{ij}}
\end{equation}
inducing a Shannon entropy reading

\begin{equation}
S_{ij}=\langle H_{ij}\rangle+\ln\zeta_{ij}=\left(\frac{\xi_{ij}}{\xi_{ij}-1}\right)-\ln[\xi_{ij}-1]+\ln m_{ij}
\end{equation}
and a FIM reading

\begin{equation}
F_{ij}=\langle(H'_{ij})^2\rangle=\frac{\xi_{ij}^2}{m^2}\left(\frac{\xi_{ij}-1}{\xi_{ij}+1}\right);
\end{equation}
the expression above holds true for the values of the parameter $\xi_{ij}$ ensuring that the Pareto distribution exists, i.e. $\xi_{ij}>1$. Besides, the convergence of the second, negative moment of the (conditional) Pareto distribution ensures that its FIM does not diverge as well.

\subsubsection{Conditional log-normal model}

The conditional log-normal model is defined by the probability distribution

\begin{equation}
q_{ij}(w_{ij}|a_{ij}=1)=\frac{e^{-\xi_{ij}\ln(w_{ij})-\gamma_{0}\ln^2(w_{ij})}}{\sqrt{\frac{\pi}{\gamma_{0}}}e^{\frac{(\xi_{ij}-1)^2}{4\gamma_{0}}}}
\end{equation}
inducing a Shannon entropy reading

\begin{equation}
S_{ij}=\langle H_{ij}\rangle+\ln\zeta_{ij}=\frac{1-\xi_{ij}}{2\gamma_0}+\frac{1}{2}\left[1+\frac{1}{2}\ln\left(\dfrac{\pi}{\gamma_0}\right)\right]
\end{equation}
and a FIM reading

\begin{equation}
F_{ij}=\langle(H'_{ij})^2\rangle=e^{\xi_{ij}/\gamma_0}(1+2\gamma_0+\xi_{ij}+\xi_{ij}^2);
\end{equation}
the expression above holds true for the values of the parameter $\gamma_0$ ensuring that the log-normal distribution exists, i.e. $\gamma_0>0$.

\end{document}